\documentclass[fleqn,usenatbib]{mnras}
\usepackage{newtxtext,newtxmath}
\usepackage[T1]{fontenc}
\DeclareRobustCommand{\VAN}[3]{#2}
\let\VANthebibliography\thebibliography
\def\thebibliography{\DeclareRobustCommand{\VAN}[3]{##3}\VANthebibliography}

\usepackage{subcaption}
\usepackage{ulem}
\usepackage{soul}
\usepackage{graphicx}	
\usepackage{amsmath}	

\usepackage{subcaption} 
\usepackage{comment}

\newcommand{\msun}{\,M$_{\odot}$}

\title[X-ray cavities in TNG-Cluster]{X-ray cavities in TNG-Cluster: AGN phenomena in the full cosmological context}

\author[M. Prunier et al.]{
Marine Prunier,$^{1,2,3}$\thanks{E-mail: marine.claude.anne.prunier@umontreal.ca}
Julie Hlavacek-Larrondo,$^{1,2}$
Annalisa Pillepich,$^{3}$
Katrin Lehle,$^{4}$ and Dylan Nelson $^{4}$
\\
$^{1}$Département de Physique, Université de Montréal, Succ. Centre-Ville, Montréal, Québec, H3C 3J7, Canada\\
$^{2}$Centre de recherche en astrophysique du Québec (CRAQ)\\
$^{3}$Max-Planck-Institut f{\"u}r Astronomie, K{\"o}nigstuhl 17, D-69117 Heidelberg, Germany\\
$^{4}$ Universit\"{a}t Heidelberg, Zentrum f\"{u}r Astronomie, ITA, Albert-Ueberle-Str. 2, 69120 Heidelberg, Germany} 

\date{Accepted XXX. Received YYY; in original form ZZZ}

\pubyear{\the\year{}}

\begin{document}
\label{firstpage}
\pagerange{\pageref{firstpage}--\pageref{lastpage}}
\maketitle

\begin{abstract}
Active galactic nuclei (AGN) feedback from supermassive black holes (SMBHs) at the centers of galaxy clusters plays a key role in regulating star formation and shaping the intracluster medium, often manifesting through prominent X-ray cavities embedded in the cluster's hot atmosphere. Here we show that X-ray cavities arise naturally due to AGN feedback in TNG-Cluster. This is a new suite of magnetohydrodynamic cosmological simulations of galaxy formation and evolution, and hence of galaxy clusters, whereby cold dark matter, baryon dynamics, galactic astrophysics, and magnetic fields are evolved together consistently. We construct mock Chandra X-ray observations of the central regions of the 352 simulated clusters at z=0 and find that $\sim$39 per cent contain X-ray cavities. Identified X-ray cavities vary in configuration with some still attached to their SMBH, while others have buoyantly risen. Their size ranges from a few to several tens of kpc. TNG-Cluster X-ray cavities are underdense compared to the surrounding halo and filled with hot gas ($\sim$10$^8$K); 25 per cent of them are surrounded by an X-ray bright and compressed rim associated with a weak shock (Mach number $\sim$1.5). Clusters exhibiting X-ray cavities are preferentially strong or weak cool-cores, are dynamically relaxed, and host SMBHs accreting at low Eddington rates. We show that TNG-Cluster X-ray cavities originate from episodic, wind-like energy injections from central AGN. Our results illustrate the existence and diversity of X-ray cavities simulated in state-of-the-art models within realistic cosmological environments and show that these can form without necessarily invoking bipolar, collimated, or relativistic jets.
\end{abstract}





\section{Introduction} \label{sec:intro}
Galaxy clusters -- hereafter clusters -- are the most massive self-gravitating systems in today's Universe, hosting hundreds to thousands of galaxies. Clusters also contain a $10^7$ to $10^8$ K hot plasma, the intracluster medium (ICM), which emits X-rays by thermal bremsstrahlung and helium-like lines from heavy elements, such as iron. This radiation cools the gas, causing it to condense and migrate gravitationally towards the cluster core, forming a cooling flow in the absence of some counteracting heating mechanism \citep{Fabian1994}. If cooling flows accumulate onto the brightest cluster galaxy (BCG), they are expected to trigger a significant level of star formation. However, observations have revealed a lower rate of star formation than expectations based on the intensity of the cooling flows \citep{McNamara1989,Fabian1992,2003Peterson,2006Peterson}. Active galactic nuclei (AGN) feedback from the supermassive black hole (SMBH) in BCGs is generally considered the most likely heating source balancing the gas cooling losses from X-ray radiation \citep{2000Fabian,2007McNamara}.

Numerous X-ray observations of cluster cores have revealed the existence of X-ray cavities, also known as bubbles, of varying sizes in their hot atmospheres. These are thought to be the primary mechanism through which AGN energy is transferred into thermal energy in the ICM. X-ray cavities have been observed in association with radio features, such as jets and lobes \citep{2012Doria,2012Gitti}, for which mechanical powers of $10^{41}-10^{45}$ erg s$^{-1}$ have been estimated \citep[e.g.,][]{2004Birzan}. However, major uncertainties remain as to the precise processes of feedback energy transfer and heating, which could result from the dissipation of sound waves and weak shocks \citep{2008MNRAS.386..278G,2018Tang,2019Bambic,2017Li}, turbulence \citep{2014Zhuravleva,2019Mohapatra}, the mixing of hot gas filling the X-ray cavities with the surrounding gas \citep{2016Yang,2017Hillel}, or cosmic ray flux originating from jets or escaping X-ray cavities and streaming through the ICM, exciting waves in the plasma that transfer their energy to the gas \citep[e.g.,][]{1991Loew,2017Rusk,2018Ehlert}.
Observationally, X-ray cavities are characterized by line-of-sight projected, prominent, X-ray surface brightness depressions on the sky, with sizes extending from a few kpc \citep[e.g., Abell 262,][]{2004ApJ...612..817B} up to a few hundreds of kpc \citep[e.g., MS0735.6+7321,][]{2005McNamara} in diameter, often exceeding the size of their host galaxies. Multiple generations of X-ray cavities on different scales can be observed in the same system \citep[e.g., NGC 5813,][]{2011ApJ...726...86R}. The discrete nature of the X-ray cavities might result either from episodic SMBH energy injections or from the fragmentation of a continuous AGN jet. These X-ray cavities are often seen in symmetric pairs around the central SMBH, supporting the hypothesis that they result from the interaction between relativistic bipolar jets and the intracluster gas \citep{2017Hillel}. Some of the X-ray cavity edges are ringed by a bright rim where the gas is compressed: apart from rare exceptions, no significant temperature jumps associated with strong shocks at X-ray cavity boundaries have been inferred, even though some systems exhibit a weak shock front with Mach numbers of $1.2-1.7$, such as Perseus \citep{2008MNRAS.386..278G}, Virgo \citep{2007ApJ...665.1057F}, and Hydra A \citep{2005ApJ...628..629N}. Observed X-ray cavities are much less dense than the ICM at the same pressure and, as they rise, their volumes increase to maintain pressure equilibrium with the surrounding gas, as interpreted from the observed correlation between X-ray cavity area and distance to the SMBH \citep{2008diehl,2016Shin}. Furthermore, based on the estimation of very basic thermodynamic properties of X-ray cavities, a significant correlation between the powers contained in X-ray cavities and the luminosity of the X-ray emitting gas in a specific radius where cooling occurs has been shown \citep{2014Panagoulia2,2022Timmerman}. This suggests that the X-ray cavities can energetically compensate for the cooling of the cluster gas due to bremsstrahlung radiation.

Despite a number of many beautiful or even spectacular examples, the observational study of X-ray cavities is challenging, because of the limited spatial and spectral resolutions of telescopes. Over the past decades, a large number have been detected by the Chandra X-ray Observatory and XMM-Newton X-ray space telescopes \citep[e.g. see][]{2007McNamara}, and several comprehensive studies in both nearby and distant massive galaxies, groups and clusters have helped to infer their underlying properties \citep{2004Birzan,2006Rafferty,2008diehl,2012Hlavacek,2014Panagoulia2,2015Hlv,2016Shin}. 
Nevertheless, the nature of these structures, their formation, and evolution in the ICM remain unclear.

In the realm of numerical simulations, numerous studies in (generally idealized) clusters set-up \citep{2001Reynold,2004Omma,2009Bruggen,2015Li,ehlert2021,2022Beckman,2024Fournier} have explored the impact of AGN jets on the ICM and shown that such an interaction is complex and non-linear. They suggest that jet-inflated X-ray cavities are formed from the expansion and displacement of the ICM by the motion of ultra-hot gas (>10$^{8-10}$ K) in a purely gas-dynamical process \citep[e.g.][]{2014Perucho}, or via cosmic rays dominated jets \citep[e.g.][]{2023Lin}, or magnetically dominated jets \citep[e.g.][]{2008Nakamura}. Resolved jet simulations demonstrate that AGN jets can generate X-ray cavities in the intracluster gas, which rise buoyantly, increasing turbulence within the cluster and driving weak shock waves and sound waves that can viscously dissipate in the ICM \citep[e.g.,][]{2004Ruszkowski,2007Bruggen,2009Sternberg}.
Even though AGNs are modeled to inject a lot of energy into the ICM (of the order of 10$^{44-45}$ erg s$^{-1}$, \citep[e.g.,][]{2017Weinberger_jet_ICM,2023Chen_Heinz} the effective mechanism for energy transmission remains poorly understood also in numerical experiments. 

Their survival is also still under scrutiny: whereas observed X-ray cavities in clusters seem to be stable and long-lived \citep[between 1 and 100 million years (Myr),][]{2004Birzan}, the ones formed purely through hydrodynamic processes, in contrast, disperse within shorter timescales due to several instabilities such as Kelvin-Helmholtz, Rayleigh-Taylor, and Richtmyer-Meshkov. However, viscosity or magnetic fields have been found with idealized simulations to stabilize these structures and possibly suppress the development of instabilities \citep{2005MNRAS.357..242R}.

Despite the numerous physical insights provided therein, simulations of idealized clusters so far have not systematically explored representative samples of halos, i.e. with varying cluster masses, assembly times, and BCG morphologies. 
Furthermore, these idealized clusters are, by construction, not influenced by their environment through gas inflows, tidal forces, and mergers. Finally, while their AGN feedback properties have been explored across large parameter spaces, the emergence of X-ray cavities and their impact on the evolving galaxy populations remain untested in the full cosmological context.

In this paper, we significantly advance our understanding by 
investigating whether and how X-ray cavities are produced in full cosmological simulations of clusters and their galaxies. In particular, we analyze the outcome of the new TNG-Cluster project \citep{2024Nelson}, which is a suite of zoom-in high-resolution magnetohydrodynamic cosmological simulations of 352 massive galaxy clusters with $M_{500\text{c}} = 10^{14.0}-10^{15.3}$ \msun \footnote{$M_{500\text{c}}$ denotes in a sphere whose mean density is 500 times the critical density of the Universe, at the time the halo is considered.} sampled from a 1 Gpc-sized cosmological box. TNG-Cluster, by using the IllustrisTNG galaxy formation model, offers a unique combination of highly-resolved and realistic high-mass galaxies and clusters, well suited to the study of the X-ray cavity population.

In fact, TNG-Cluster differs in fundamental ways from the simulations used so far to understand X-ray cavities: the clusters therein are not idealized but are evolved across billions of years of cosmic evolution together with their central and satellite galaxies (and their gaseous and stellar components) and together with their evolving SMBHs, cosmological gas accretion, and mergers. Previous studies within the IllustrisTNG framework have demonstrated that TNG50 Milky Way-like galaxies display eROSITA-like bubbles associated with kinetic injections from SMBH feedback \citep{2021PillepichErosita} and that Perseus-like cluster cores from TNG-Cluster exhibit disturbed X-ray morphologies such as ripples, X-ray cavities, and shock fronts that closely resemble observations \citep{2024Truong}. Independently from IllustrisTNG, alternative implementations of SMBH-driven feedback in cosmological setups have been shown to produce features similar to observed X-ray cavities in a simulated cluster \citep[RomulusC, via thermal energy injections with shut-off cooling,][]{2019Tremmel} and, very recently, in a set of simulated galaxy groups \citep[the Hyenas, based on the bipolar jet-like model of SIMBA,][]{2024Jennings}. This provides further incentive to investigate the emergence of X-ray cavities in the full cosmological context, across halo mass scales and galaxy formation models.

Here, we present the first paper in a series that aims to study X-ray cavities with the TNG-Cluster simulation suite. In particular, in the following, we quantify their global and spatially-resolved morphologies and demographics, by offering an overview of the diversity of X-ray cavities that are naturally produced in TNG-Cluster. To this end, we produce and analyze X-ray Chandra mock images of the central region of all clusters at redshift 0, $z=0$, in the simulation. In a second paper (Prunier et al. in prep), we will carry out a comprehensive and apples-to-apples comparison between real X-ray cavities identified in a volume-limited sample of observed clusters with an analog subset of TNG-Cluster halos, by quantifying the characteristics and energetics of the two X-ray cavity populations. While this first paper will mostly utilize physical quantities directly inferred from the simulation output, the second paper will aim to replicate, as closely as possible, the methods used in an observational study, allowing us to assess how well the simulation results align with real observations.

This paper is organized as follows. In Section \ref{sec:meth}, we detail the TNG-Cluster simulation and the methods used to identify and characterize X-ray cavities. Section \ref{sec:results} presents our key findings, highlighting the morphological diversity of TNG X-ray cavities and providing statistics and demographics of the population across clusters. In Section \ref{sec:origin}, we explore the connection between TNG SMBH feedback and the formation of X-ray cavities. In Section \ref{sec:discussion}, we interpret our results and their implications before concluding in Section \ref{sec:conclusion}.

\section{Methodology: Simulated Clusters and their X-ray Mock Observations}
\label{sec:meth}

\subsection{The TNG-Cluster simulation suite}
\label{subsec:tng-cluster}

TNG-Cluster\footnote{\url{https://www.tng-project.org/cluster/}} \citep{2024Nelson} is a new spin-off project of IllustrisTNG \citep[TNG hereafter,][]{2019NelsonPublicReleaseTNG}. It is a 1 Gpc addition to the TNG simulation suite (with flagship runs known as TNG50, TNG100, TNG300). It comprises a series of 352 zoom-in simulations of high-mass galaxy clusters (M$_{\text{cluster, 500c}} \geq 10^{14}$ \msun) with a baryonic mass resolution of about $10^{7}$~\msun. 
TNG-Cluster, likewise TNG, employs an adaptive mesh refinement scheme for the hydrodynamics, adjusting cell sizes based on environmental density. In TNG-Cluster, the spatial resolution can be as small as 50 parsecs in the highest-density regions, enabling the detailed simulation of small-scale structure\footnote{Within 50 kpc of cluster cores, the median resolution of gas cells is 2.3 kpc (1st percentile: 1.2 kpc, 99th percentile: 2.9 kpc). Within the innermost 10 kpc of the clusters, the median resolution is 1.7 kpc (1st percentile: 1.4 kpc, 99th percentile: 2.0 kpc).}. 
The project aims to create a sample of clusters that match the mass-redshift space of large cluster surveys, such as the ROSAT MCXC meta-catalog \citep{2011A&A...534A.109P}, with hundreds of simulated massive clusters across cosmic epochs. TNG-Cluster adopts the fiducial TNG cosmology, consistent with Planck 2016 \citep{2016A&A...594A..13P}, with parameters $h = 0.6774$, $\Omega_m = 0.3089$, $\Omega_{\Lambda} = 0.6911$, $\Omega_b = 0.0486$, $\sigma_8 = 0.8159$, and $n_s = 0.9667$. 

The physics of TNG-Cluster is based entirely on the model presented in \citep{2018MNRAS.473.4077P,2017RainerW} and implemented in TNG. It is a comprehensive and validated physical model of the formation and evolution of galaxies and clusters, including gas heating and cooling, star formation, evolution of stellar populations and chemical enrichment, stellar feedback, as well as accretion, multimode feedback, and merging of SMBHs. TNG-Cluster, thus, enables a detailed study of the ICM at the heart of clusters, and its interaction with AGN feedback.

\paragraph*{SMBH feedback in TNG-Cluster}
In the TNG model, and hence in TNG-Cluster, feedback from SMBHs is injected into the surrounding medium in the form of thermal or kinetic energy, for high and low accretion rates respectively \citep{2017RainerW}, in addition to a radiative feedback \citep{2013Vogelsberger}. The energy injected in feedback is a fraction of the energy available through mass growth via gas accretion (in turn, via a Bondi formula). SMBHs transition from thermal (also called quasar feedback) to kinetic feedback (also called mechanical or wind feedback) depending on their instantaneous accretion rate and mass. 

All the primary 352 BCGs of TNG-Cluster are in kinetic mode feedback at $z=0$  \citep[having transitioned for the first time from thermal to kinetic mode between redshift 6 and 2,][]{Rohr2024_cool}: their central SMBHs inject kinetic energy by inputting a unidirectional momentum kick. The direction of each kick is randomly chosen and varies between each injection episode. In some sense, this choice mimics a jet reorientation, but with each outburst delivering energy in a single event rather than over tens of millions of years, as seen in some idealized simulations \citep[e.g.,][]{2018Cielo}. Over time, however, the randomness of the injection directionality leads to an isotropic energy distribution. The feedback energy released is parametrized as \begin{math} \Delta \dot{E}_{\text{kin}} = \epsilon \dot{M}_{\text{SMBH}} c^2 \end{math}, where the maximum coupling efficiency $\epsilon$ is 0.2 \citep{2017RainerW,2018MNRAS.473.4077P}. The frequency of kinetic feedback is determined by the minimum energy that must be accumulated in the kinetic accretion mode by the SMBH before feedback is released. In TNG-Cluster, where the timing of kinetic energy injections has been recorded for each SMBH, this frequency around $z=0$ corresponds to, typically, 1 injection every 10-100 Myr. 

As evinced from this description, by construction, in these simulations there are no bipolar, collimated, SBMH-driven jets akin to those implemented in idealized cluster simulations and to those identified in e.g., radio galaxies \citep[e.g.,][]{2004ApJ...612..817B, 2022Timmerman}. Also, no cosmic ray physics is implemented \citep[although see][]{2024Ramesh}. Rather, the SMBH kinetic feedback included in TNG-Cluster is a subgrid model meant to mimic (the effects of) high-velocity 
accretion-disk winds or small-scale jets from low-luminosity SMBHs \citep{2014Yuan} and designed primarily to halt star formation in massive galaxies. Such a model has been shown to be able to drive high-velocity outflows of hot gas \citep{2019Nelson_TNG50} with varying large-scale ($\gtrsim 10$ kpc) manifestations in galaxies of different types and masses, from Milky Way analogs \citep{2021PillepichErosita,2023arXivPillepich_TNG50MW_Like,2023aRamesh} to the BCGs at the center of Perseus-like clusters \citep{2024Truong}. In the following, we further test this model in terms of its ability (or not) to create X-ray cavities in clusters and its interaction with the ICM.

\begin{figure*}
    \centering
    \includegraphics[trim=9cm 8cm 0cm 10cm, clip, width=\paperwidth]{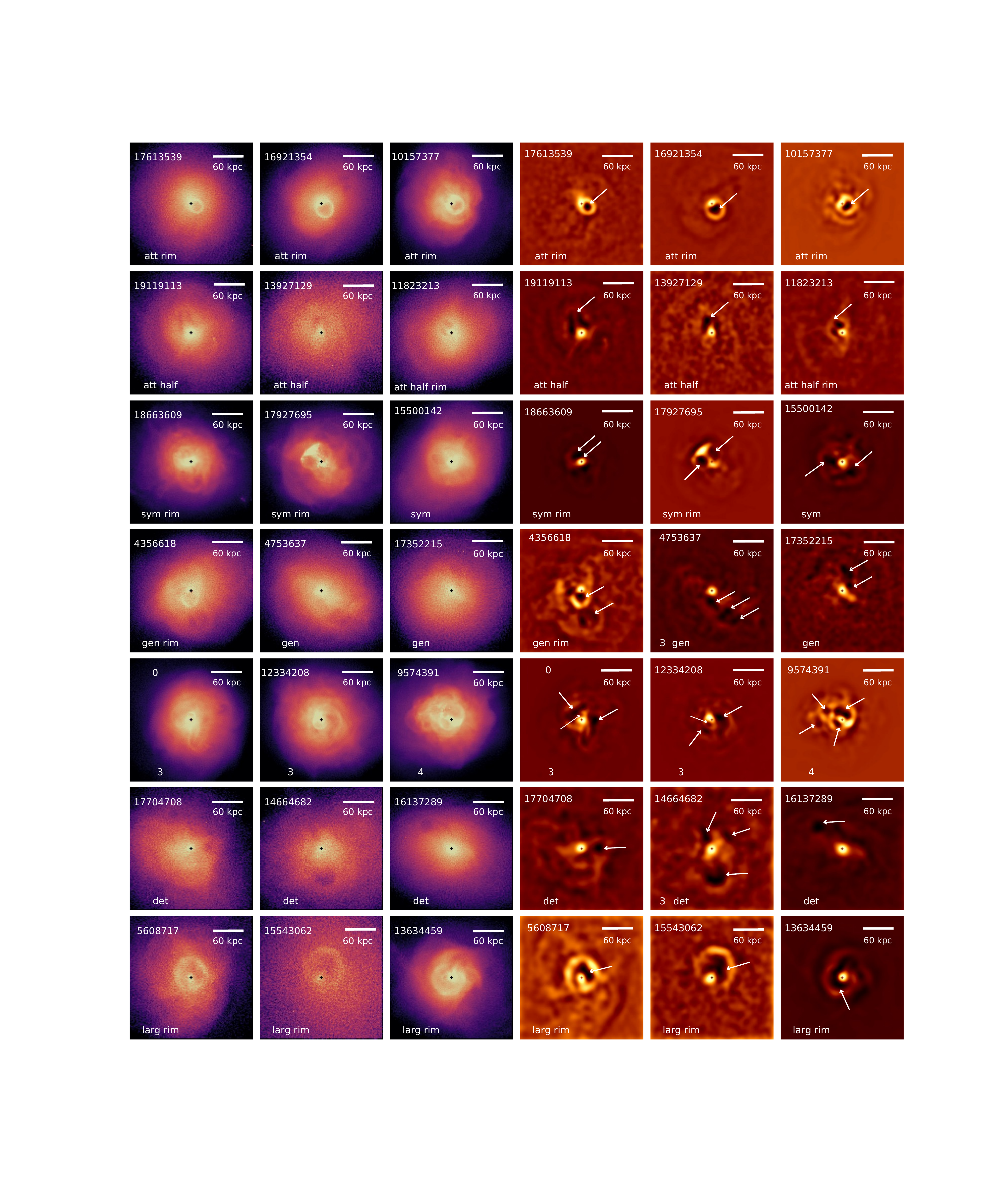}
    \vspace{-0.9cm}
    \caption{Gallery of TNG-Cluster BCGs with visible X-ray cavities at $z=0$. On the \textit{Left columns:} we show Mock Chandra surface brightness images of the simulated clusters; on the \textit{Right columns:} X-ray maps processed with an unsharp mask filter to highlight X-ray-depleted regions. 
    The black crosses denote the position of the SMBH. White arrows point out the X-ray cavity locations. Letters at the bottom right are flags characterizing the cluster's X-ray cavities:\textit{ att}- attached to the SMBH, \textit{rim}- presence of a bright rim,\textit{ half}- one X-ray cavity cut in half by a bright gaseous emission, \textit{sym}- pair of axisymmetric X-ray cavities, \textit{gen}- several generations, \textit{3 or 4}- more than two identified X-ray cavities in the same cluster \textit{det}- detached and rising in the ICM, \textit{larg}- very large surface-brightness depressions surrounding the SMBH.
    This gallery highlights the diversity of X-ray cavity types, morphologies, and evolutionary stages produced in TNG-Cluster.}
    \label{fig:panel_cav}
\end{figure*}

\subsection{Mock Chandra X-ray images} 
\label{subsec:mocks}
As the observational study of X-ray cavities is prone to biases, influenced by data quality and instrumental effects, we include a layer of observational realism for their identification in simulated clusters. Our methodology involves generating simulated Chandra images, or ``mock'', rather than relying solely on ideal X-ray maps extracted directly from the simulations. The Chandra X-ray imaging instrument ACIS-I, which offers high spatial resolution (pixel resolution 0.492 arcsec) and large field of view (16.9 arcmin), is frequently used for the study of X-ray cavities. To investigate these structures in TNG-Cluster, we build mock ACIS-I Chandra X-ray images (field of view of 2$\times r_{500}$) of the 352 primary-zoom halos at $z=0$ within the simulation. Each of these mock images is centered at the location of the SMBH of the halo's BCG\footnote{ Throughout this work, by BCG we denote the most massive galaxy of the (friends-of-friends halo corresponding to each) cluster. The central SMBH of each cluster should be the most massive SMBH that is gravitationally bound to the BCG at $z=0$.}.

We use pyXSIM \citep{2016ascl.soft08002Z} and the SOXS \citep{2023ascl.soft01024Z} software suite for simulating X-ray photons and producing mock observations. Photons in the energy band of 0.5–10.0 keV are generated for each cluster over a cubic region of $\pm1\times$r$_{500}$ around the central SMBH. All gas cells in such cubic regions are considered for each cluster, i.e. not just gas parcels that are gravitationally-bound to the BCG or friends-of-friends cells. We exclude only star-forming gas, i.e. cells with positive cooling rates (i.e. net heating), and those low-resolution cells from outside the zoom-in, high-resolution, re-simulation regions, that happen to be in our cubic volume selection at the time of analysis. For each gas cell, a mock spectrum is generated based on its gas density, temperature, and metallicity, assuming the simulation abundance ratios, a single-temperature APEC model \citep{2001ApJ...556L..91S}, and galactic absorption with a hydrogen column density of 4$\times$10$^{20}$cm$^{-2}$. The spectra of all gas cells in the adopted core region are summed up together. This produces a large initial random sample of (``intrinsic'') photons for each cluster and each sight line, which is later used by the SOXS instrument simulator to draw subsamples of photons to create “observed” X-ray events. 

More specifically, mock Chandra ACIS event files are produced by projecting the photons onto a detector plane, and convolving them with an instrument model for the ACIS-I detector -- we use instrumental responses files of Cycle 19, and with the (on-axis) point-spread function of ACIS. We limit the energy band of 0.5–7.0 keV to match the ACIS broad energy band. To mimic as closely as possible real observational data, we also include X-ray emission of satellite galaxies belonging to the same halo in the image, as well as instrumental and cosmic X-ray backgrounds and the galactic Milky Way foreground. The mock X-ray observations are created for each simulated cluster in a random orientation (the z-axis of the TNG-Cluster box). As a sufficient number of X-ray photons is required to identify X-ray cavities, we chose an exposure time of 200 kilo-seconds (ks) for our images. This integration time makes it possible to visualize fine features in cluster cores and aligns with the typical total exposure time of well-studied clusters harboring X-ray cavities by Chandra \citep[e.g., NGC 4636,][]{2009ApJ...707.1034B}. 
All clusters are positioned at a fixed angular distance of 200 Mpc from ACIS-I, which corresponds to a redshift of 0.05, resulting in a resolution of $\sim$0.5 kpc with the instrument specifications. The final images are background-subtracted, and point sources are removed using the \texttt{CIAO} wavelet detection method and \texttt{dmfilth} to replace source pixels.

\subsection{Identification of X-ray cavities}
\label{sec:um}
We identify X-ray cavities in the simulated clusters by employing unsharp masking (UM) on the mock Chandra X-ray observations obtained and described above. UM is a spatial frequency filtering technique that enhances contrast and reveals small-scale inhomogeneities in an image. This method, widely used in previous studies of X-ray cavities \citep[e.g.,][]{2014Panagoulia2,2016Shin}, involves convolving Chandra mock images with Gaussian kernels of varying sizes, e.g., 6, 8, 12, and 16 pixels (equivalent to 3'', 4'', 6'', 8'' or about 3, 4, 6, 8 kpc, respectively), and subsequently subtracting the images smoothed with larger kernels (respectively 12 and 16 pixels) from those smoothed with smaller kernels (6 and 8 pixels). Having filters of different sizes enables better visualization of small and large X-ray cavities that can coexist in the same cluster. We identify X-ray cavities in TNG-Cluster halos by visually inspecting both the Chandra-mock and two unsharp masked images for each cluster. We select only X-ray cavities that are clearly visible in both the original mock image and in the UM images or that are unambiguously visible in the UM images.

\subsection{Sizes of X-ray cavities and other properties}
\label{sec:props}
We manually estimate the physical size of each X-ray cavity as typically done in observations, i.e. by superimposing an ellipse at the location of the X-ray cavity on both the original mock Chandra X-ray map and its UM counterparts. As in the observational studies, we systematically assume X-ray cavities to be prolate ellipsoids, with axes parallel and perpendicular to the line that connects the SMBH to the center of the X-ray cavity \citep[e.g.][]{2010Dong}.
We only measure the projected 2D quantities:  even though we could estimate the actual 3D cavity sizes and volumes in the simulation, throughout we prefer to adopt an observer's perspective on the simulated data. We leave for future work the estimation of the 3D morphology of the X-ray cavities and the bias it may introduce in the calculation of the energetics.
As the size of each X-ray cavity, we take the mean radius i.e. the average of the semi-major and semi-minor axes in kpc. The projected area is computed as \begin{math}\pi r_a r_b \end{math}.
While this method may lack precision, it reflects the current practice among observers who do not typically use automated methods for detecting and measuring X-ray cavity sizes \citep[although pioneering efforts have been made recently in this direction with][]{Plsek}. Due to the simplifying assumptions of ellipsoidal shape and symmetry, as well as projection effects i.e. X-ray cavity orientation, errors on X-ray cavity sizes are large: we assume an uncertainty of the order of $ 20$ per cent \citep[e.g.,][]{2015Hlv}.

\begin{figure}
    \centering
    \includegraphics[width=0.475\textwidth]{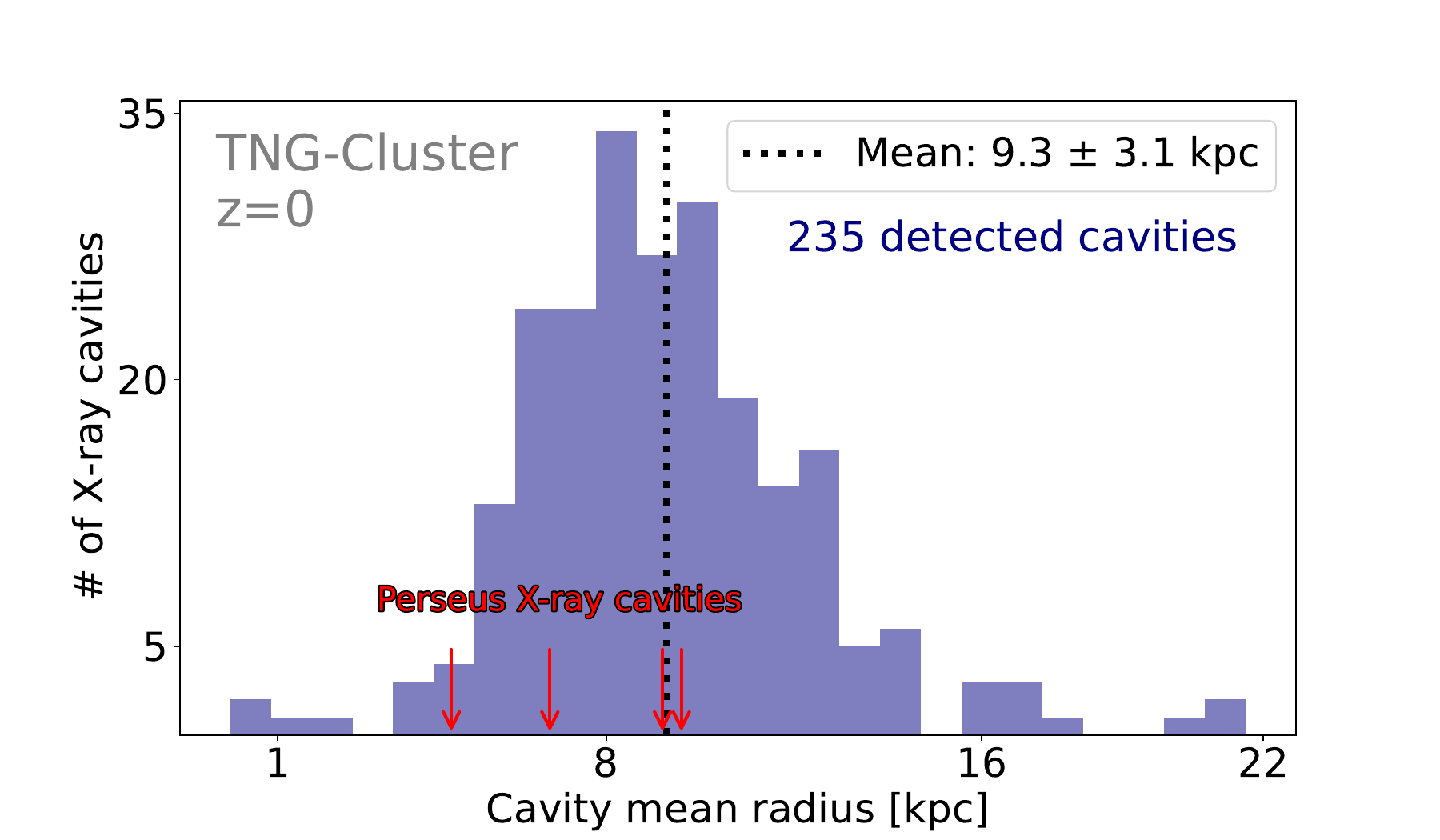}
    \caption{Distribution of TNG-Cluster X-ray cavity sizes (i.e. mean radius; average of the axes of the ellipse} in kpc) at $z=0$. Their sizes range from a few kpc to a few tens of kpc with a mean of $9.3\pm 3.1$ kpc (i.e. $\pm1$-$\sigma$ assuming a Gaussian distribution). For comparison, we also display data from X-ray cavities observed in the Perseus cluster \citep{2002MNRAS.331..369F,2000Fabian}.
    \label{fig:radius}
\end{figure}

\section{Results: A population of X-ray cavities in TNG-Cluster}\label{sec:results}

\subsection{Diversity and morphology} \label{subsec:morpho}
X-ray cavities are common in TNG-Cluster: more specifically, we identify one or more X-ray cavities in 39 per cent (136) of the 352 clusters in the simulation, for a total number of 233 X-ray cavities. We emphasize here that this analysis is conducted at a single snapshot, corresponding to redshift 0, capturing the occurrence of X-ray cavities only at the present epoch. In the panel of Figure~\ref{fig:panel_cav}, we present a collection of TNG-Cluster halos at $z=0$ with X-ray cavities visible as X-ray surface brightness depressions. The left columns display mock Chandra X-ray observations of the cluster's core ($240\times240$ kpc), while the right columns show these images processed with an unsharp mask filter to contrast the X-ray-depleted regions (see Section~\ref{sec:um}).

We include examples of X-ray cavities with different morphologies and at different stages of evolution:  from the defined X-ray cavity of the cluster in the top left map (TNG-Cluster ID 17613539), likely still undergoing inflation, to the fainter, rising one in the system with ID 14664682. We showcase a variety of configurations, including single, pairs (e.g., TNG-Cluster ID 155500142), and multiple X-ray cavities (e.g., TNG-Cluster ID 0), attached or detached from the SMBH. Notably, we also identify multiple generations of X-ray cavities within the same image (e.g., TNG-Cluster ID 4753637).

To organize the description of these X-ray cavities and qualify their diversity in the TNG-Cluster simulation, we define eight flags. These flags are assigned purely based on visual inspection to attempt to qualify the complexity of the X-ray cavity population and to distinguish different configurations identified in the simulation: 
\begin{itemize}
\item \textit{att}- X-ray cavities attached to the SMBH 
\item \textit{rim}- presence of a bright rim
\item \textit{half}- one X-ray cavity cut in half by a bright emission.
\item \textit{sym}- pair of axisymmetric X-ray cavities
\item \textit{gen}- several generations of X-ray cavities in the same cluster
\item \textit{3} and \textit{4}- more than two identified X-ray cavities in the same cluster 
\item \textit{det}- detached and rising in the ICM 
\item \textit{larg}- unusually large X-ray brightness depression surrounding the SMBH
\end{itemize} 

Overall, Figure~\ref{fig:panel_cav} showcases that the properties and shapes of TNG-Cluster X-ray cavities are diverse. Attached X-ray cavities often have X-ray bright edges (flag \textit{rim}) whereas detached ones generally do not. We also see some attached X-ray cavities cut in half by a bright emission (flag \textit{half}). Moreover, TNG-Cluster X-ray cavities tend to be more regular and spherical when smaller and attached to the SMBH (e.g., TNG-Cluster ID 17613539). X-ray cavities attached or recently detached can also be preferentially elongated towards the SMBH (e.g., TNG-Cluster ID 19119113). Larger and more distant ones exhibit more deformations, for example, the system of TNG-Cluster ID 17352215.

\begin{figure*}
    \centering
    \includegraphics[trim=1.5cm 8cm 11cm 4cm, clip, width=0.6\paperwidth]{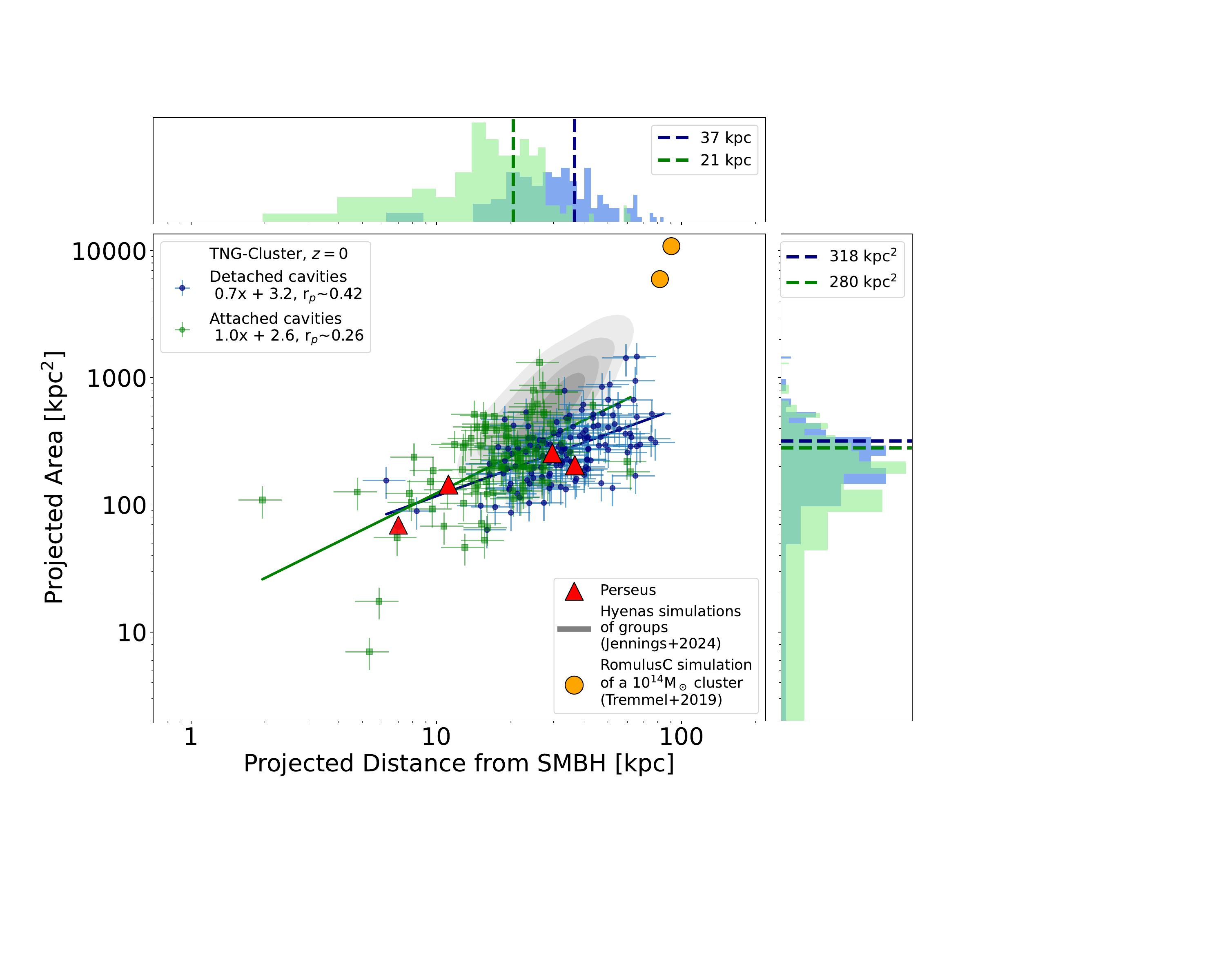}
    \caption{Area of TNG-Cluster X-ray cavities as a function of their distance from the SMBH of the BCG. Green (blue) markers represent X-ray cavities that are attached to (detached from) their SMBH. The Pearson coefficient ($r_p$) quantifies the correlation between area and distance \citep{Fisher1994}. The green and blue lines depict the best logarithmic fit for the attached and detached X-ray cavities, respectively. The histograms display the distribution of distances and areas, with dashed lines indicating the medians. For comparison, X-ray cavity data from the Perseus cluster \citep{2002MNRAS.331..369F,2000Fabian}, with a total mass of $M_{500} \sim 10^{14.5}$ \msun \hspace{0.1cm} \citep{2011Sci...331.1576S}, are included. TNG-Cluster X-ray cavities are within a comparable size range to observed ones. We also include, to provide a first-order comparison to be expanded in future works, X-ray cavities from the cosmological simulation suite of zoom-in galaxy groups Hyenas \citep[grey contours, encompassing 160 X-ray cavities in groups examined over a period of 700 Myr,][]{2024Jennings} and two X-ray cavities visible in the RomulusC low-mass cluster \citep[orange filled circles,][]{2019Tremmel}. While the Hyenas cavities are found in groups i.e. in a lower mass range than clusters, and hence are not directly comparable to this work, they are displayed here strictly for reference. The comparison between our results and those of Hyenas and RomulusC is further discussed in section \ref{sec:discussion_comp}.}
    \label{fig:area_distance}
\end{figure*}

\subsection{Spatial extent and sizes}\label{subsec:sizes}
 
The sizes of the TNG-Cluster X-ray cavities span from a few kpc to a few tens of kpc, with a mean radius across the population of $9.3$~kpc. This is shown in Figure~\ref{fig:radius}, where we depict the distribution of the mean radius of TNG-Cluster X-ray cavities measured as described in Section~\ref{sec:props}.
Amid the possibly-large uncertainties in such size estimates, TNG-Cluster sizes align with the size range reported in observational studies of clusters harboring X-ray cavities \citep[e.g.,][]{2014Panagoulia2}. A more direct and comprehensive comparison between the sizes of real and simulated X-ray cavities in the TNG model will be the focus of a future paper.

We also explore the correlation between the 2D projected area of X-ray cavities and distance to the SMBH. Our cluster mock images are generated with random orientations, therefore we do not measure the physical but the projected quantities, as in real observations. We approximate the X-ray cavity center as the ellipse center and measure its projected distance from the SMBH. Our analysis, shown in Figure~\ref{fig:area_distance}, reveals that, in TNG-Cluster and similarly to what inferred from observations (here shown for the case of Perseus), distant X-ray cavities have larger areas, suggesting a volume increase as they rise. We measure a moderate positive linear correlation between the distance from the SMBH and the area of X-ray cavities, with a Pearson correlation coefficient \( r_p \) of 0.26 for attached X-ray cavities and of 0.42 for buoyant ones. Their projected distances range from a few kpc up to 100 kpc, with a mean of 26 kpc. This result implies that in TNG these structures seem reasonably resilient to disruption processes and/or travel at high velocities. 

\begin{figure*}
    \centering
    \includegraphics[width=\textwidth]{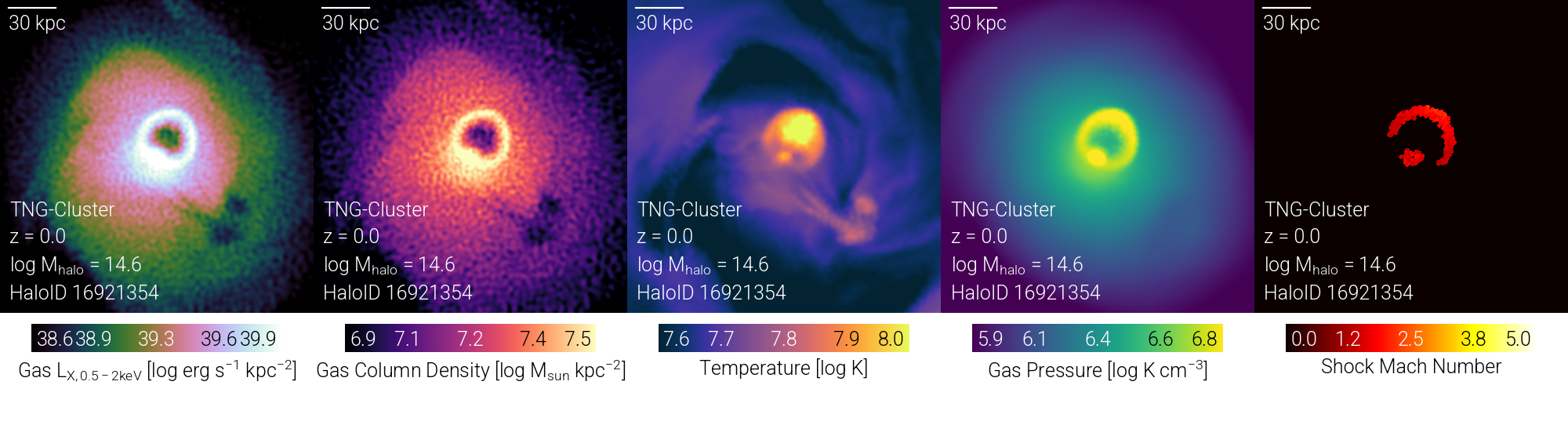}
    \caption{Maps of the thermodynamic and kinematic properties of the gas within an example TNG-Cluster system (Halo 16921354; maps side length: 200 kpc, depth: 20 kpc). Two X-ray cavities are visible: the first is attached to the central SMBH and exhibits a bright rim, while the second is detached and rising in the ICM. These X-ray cavities appear as under-dense and under-X-ray luminous regions filled with hot gas. Additionally, the attached X-ray cavity has an over-pressurized edge and a weak shock front with a Mach number of $\sim$ 1.3 (mass-weighted average Mach number along the line of sight, according to the shock finder: here the averages are obtained by only considering gas cells with shock Mach number larger than 0.9), while the detached X-ray cavity does not display such a feature. This panel illustrates the varying thermodynamic states of the gas within and surrounding the X-ray cavities in TNG-Cluster, which assumes an ideal adiabatic index for a diatomic gas ($\gamma$ = 5/3) to model the gas's thermodynamics.} It highlights the typical properties of the X-ray cavity population identified in the simulation.

    \label{fig:panel_cav_phys_pro}
\end{figure*}

\subsection{Gas properties}
\label{subsec:gas}
Figure~\ref{fig:panel_cav_phys_pro} showcases various gas properties of the core region of an example TNG-Cluster halo (ID 16921354): X-ray surface brightness, gas density, mass-weighted temperature, and mass-weighted pressure, from left to right. We also show a shock map that displays gas cells with mass-weighted Mach number > 0.9 in the map, using the cosmological shock ﬁnder implemented in the TNG model \citep{2016MNRAS.461.4441S}. We show this cluster as a representative example because it features two distinct X-ray cavities at different evolutionary stages. 

The first X-ray cavity is connected to the central SMBH, exhibiting a prominent, nearly spherical shape with a bright rim and extending $\sim$12 kpc in radius. The second X-ray cavity is rising in the ICM, exhibiting a somewhat lesser X-ray contrast compared to the surrounding medium, and lacking a distinct bright rim. Both X-ray cavities are identifiable by a $\sim$ 30-40 per cent decrease in X-ray emission compared to the surrounding gas, as inferred on the map by comparing the X-ray emission within the cavities to the ICM values within the core. As we see later, these values are typical of the TNG-Cluster population and their cavities. These X-ray cavities appear as clearly defined regions with lower density and pressure than the rest of the ICM. Moreover, the temperature map reveals that they are filled with gas of high temperature ($\sim 10^8$ K) which is 2-5 times higher than the surrounding intra-cluster gas. The X-ray-bright rim of the first X-ray cavity reaches 10$^{39.9}$ erg s$^{-1}$ kpc$^{-2}$, it is compressed and associated with a weak shock of average Mach number $\sim$ 1.4. On the other hand, the other X-ray cavity, which appears rising outwards, lacks such a shock signature. 

The examples of Figure~\ref{fig:panel_cav_phys_pro} offer insights into the typical properties of the gas observed across the X-ray cavity population of TNG-Cluster. In fact, by visually inspecting similar maps for each central region of all 352 TNG-Cluster halos at $z=0$, we identify recurring patterns associated with the presence of X-ray cavities. More specifically:

\begin{itemize}
    \item[$\blacksquare$] All TNG-Cluster X-ray cavities manifest themselves in the X-ray maps as lower-emitting regions filled with hot gas with temperatures varying between $10^{7.8}$ and $10^{8.1}$ K (10th-90th percentiles across the X-ray cavity sample).\\
    \item[$\blacksquare$] The X-ray cavity regions typically exhibit a lower density and X-ray emission with respect to the surrounding gas (by 27-48 per cent, 10th-90th percentiles). The interior of the X-ray cavities shows no signs of over- or under-pressure, implying that TNG-Cluster X-ray cavities are in approximate pressure equilibrium with respect to the surrounding ICM.\\
    \item[$\blacksquare$] About $\sim$ 25 per cent of the TNG-Cluster X-ray cavities (58 of 233) display bright and overpressurized dense edges, often coinciding with weak shocks. This, in particular, is the case for 20 X-ray cavities (9 per cent of the whole X-ray cavity sample), with Mach numbers varying between $1.2$ and $2.1$ (10th-90th percentiles). Edges seem to have similar temperatures than the gas inside the X-ray cavity.\\
    \item[$\blacksquare$] Conversely, X-ray cavities lacking bright edges or detached from the central SMBH are not associated with shocks.\\
    \item[$\blacksquare$] The ICM surrounding X-ray cavities frequently exhibits spherical ripples and pressure waves. In most cases, they are not associated with strong shocks, but sometimes with weak shocks.
\end{itemize}

\begin{figure*}
  \centering
   \hspace{-0.8cm}
  \begin{subfigure}[b]{0.55\textwidth}
    \includegraphics[width=\textwidth]{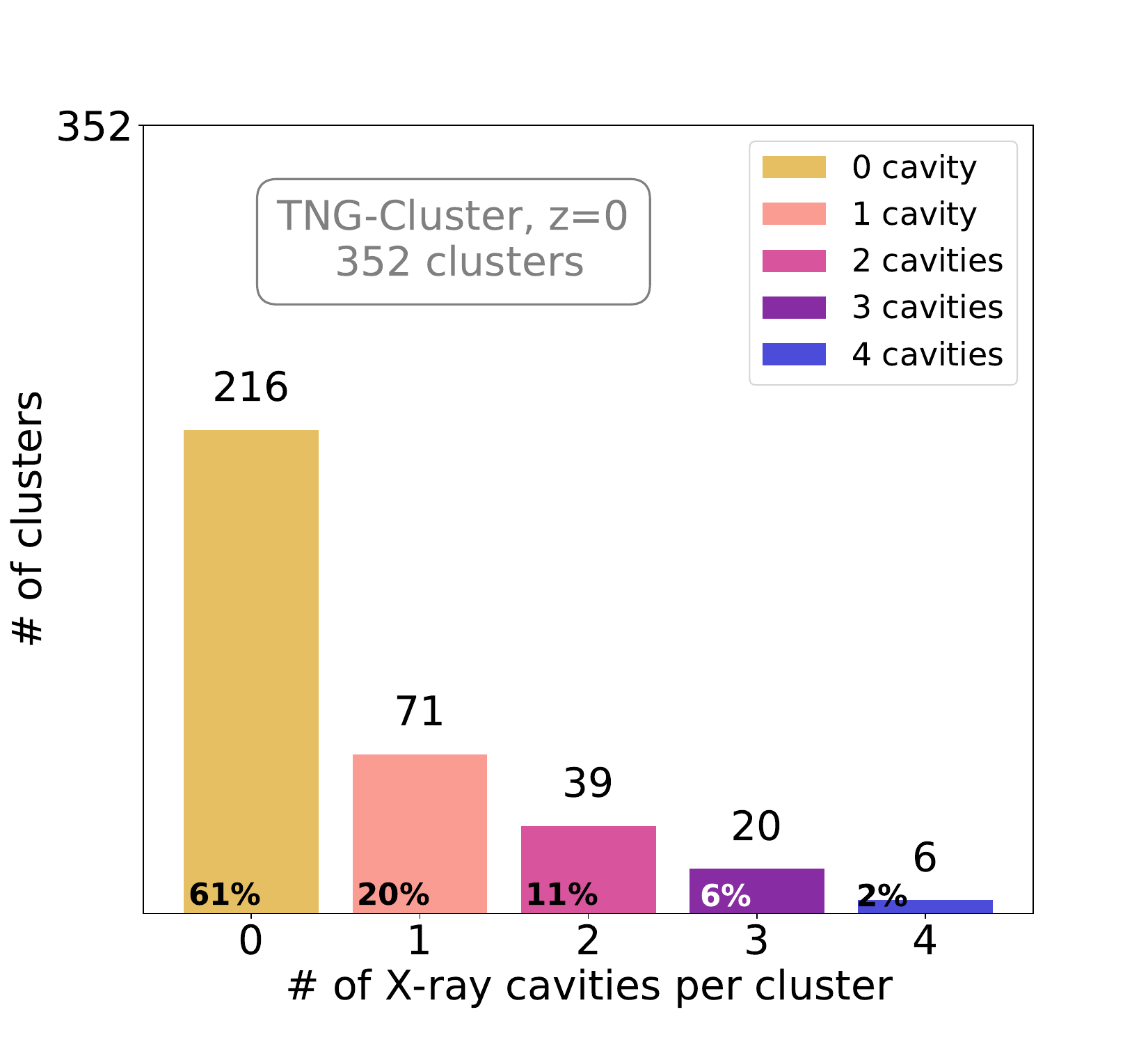}
  \end{subfigure}
  \begin{subfigure}[b]{0.3\textwidth}
   \hspace{.6cm}
   \vspace{1cm}
    \includegraphics[width=1.2\linewidth, height=1.2\linewidth]{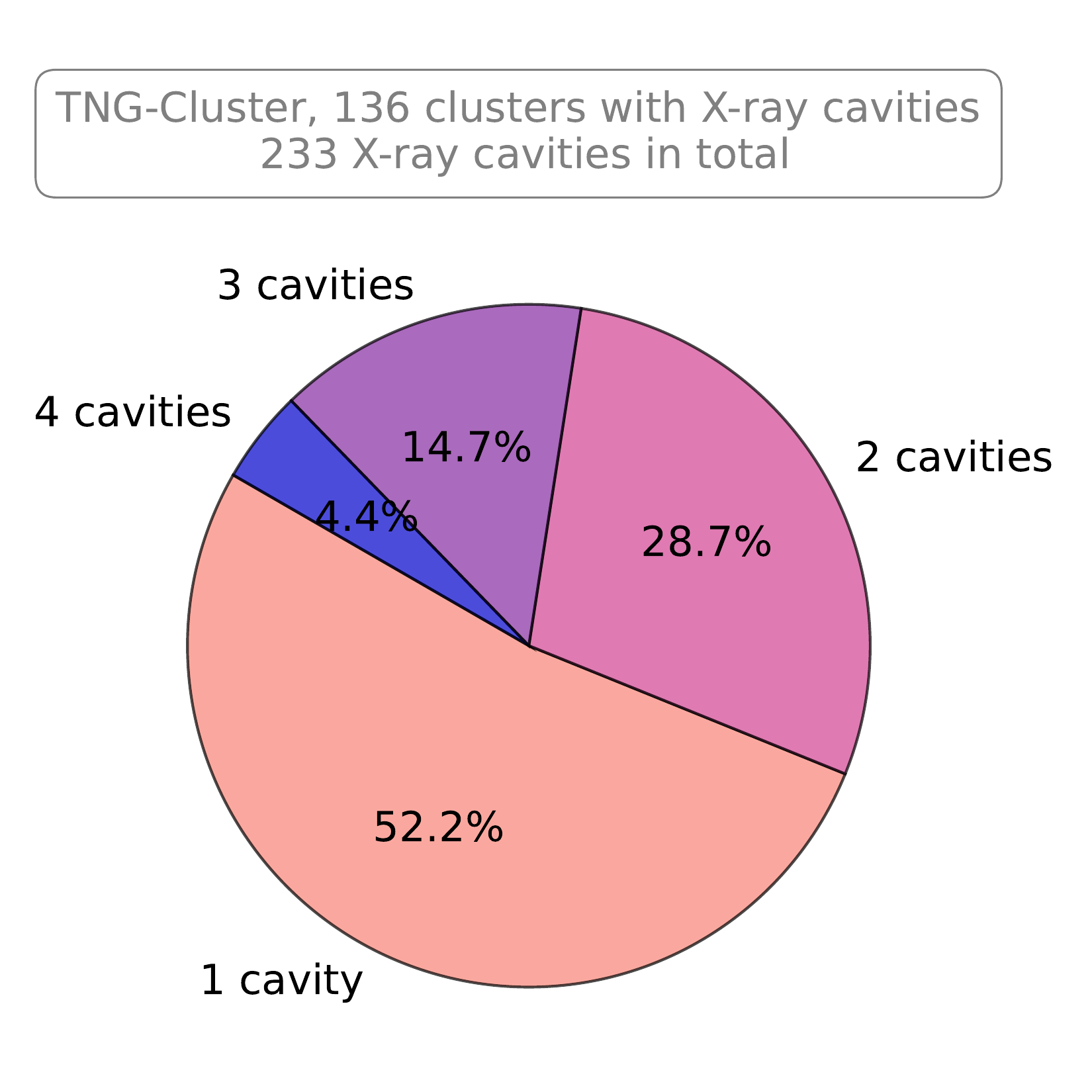}
  \end{subfigure}
  
  \vspace{-0.1cm} 
  
  \begin{subfigure}[b]{0.35\textwidth}
    \centering
    \hspace{-0.5cm}
    \includegraphics[width=\textwidth]{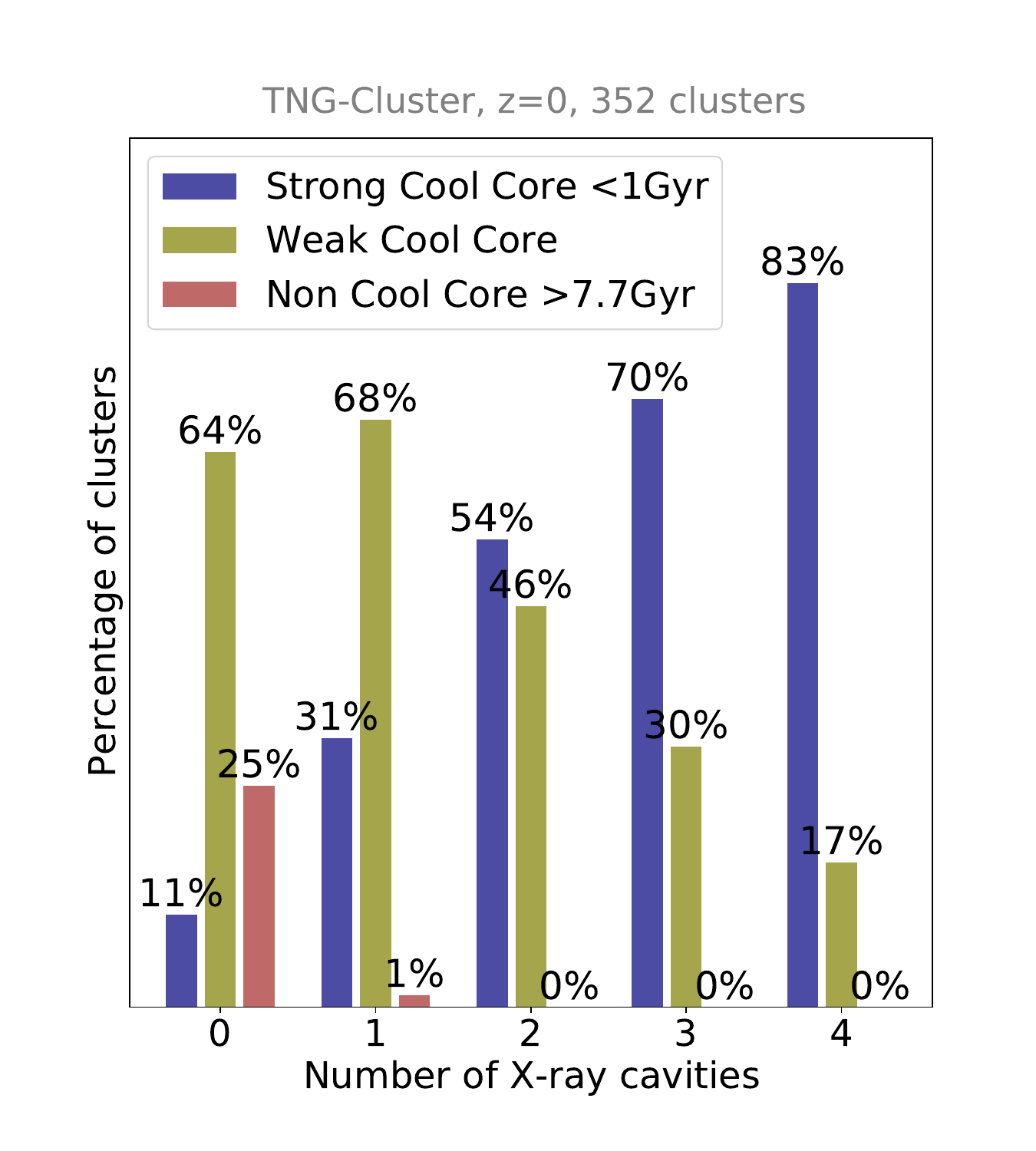}
  \end{subfigure}
  \hfill
  \begin{subfigure}[b]{0.33\textwidth}
    \centering
    \hspace{-2.5cm}
    \vspace{-0.2cm}
    \includegraphics[width=\textwidth]{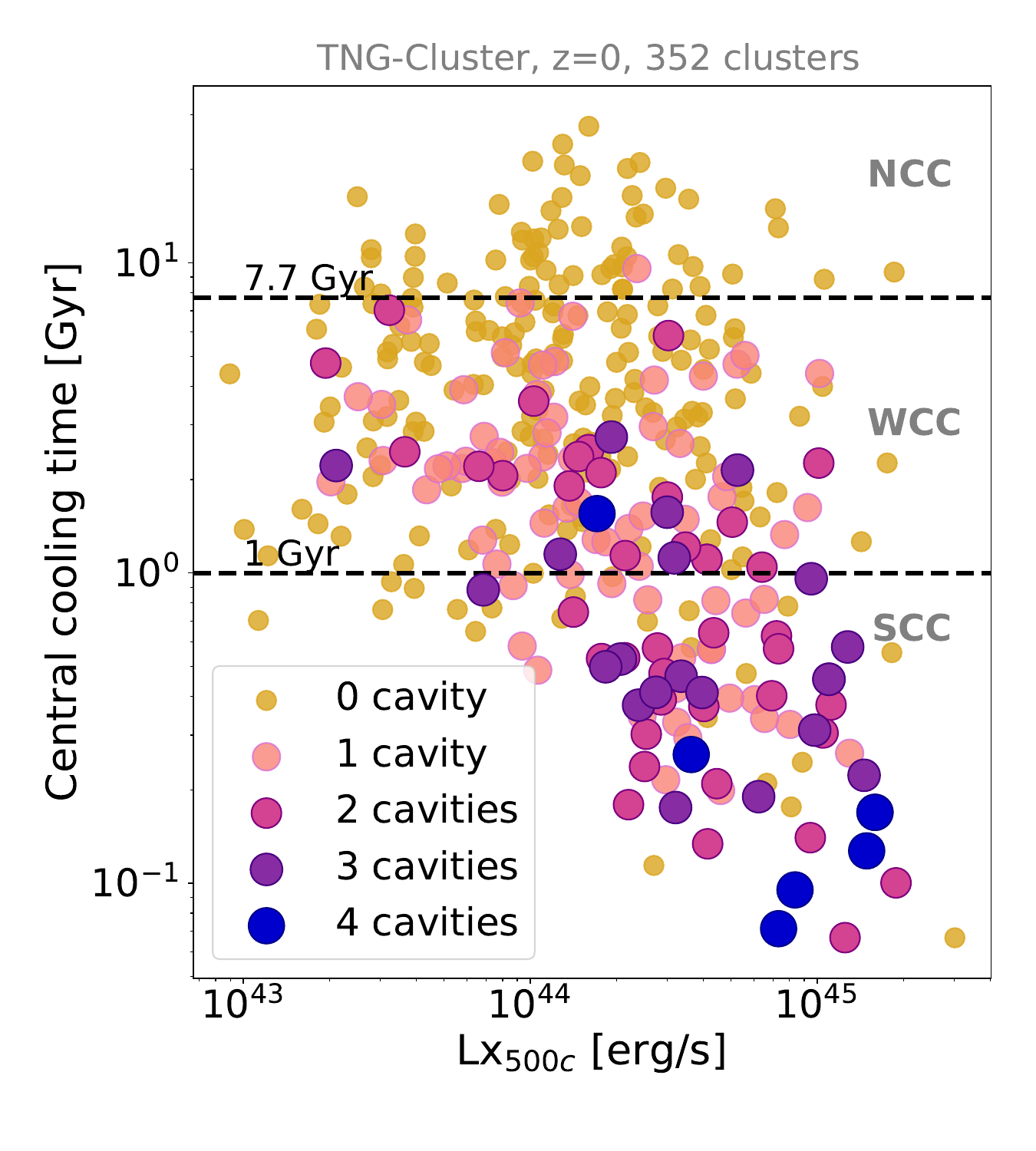}
  \end{subfigure}
    \hspace{-1cm}
  \begin{subfigure}[b]{0.3218\textwidth}
    \vspace{-4cm}
    \includegraphics[width=\textwidth]{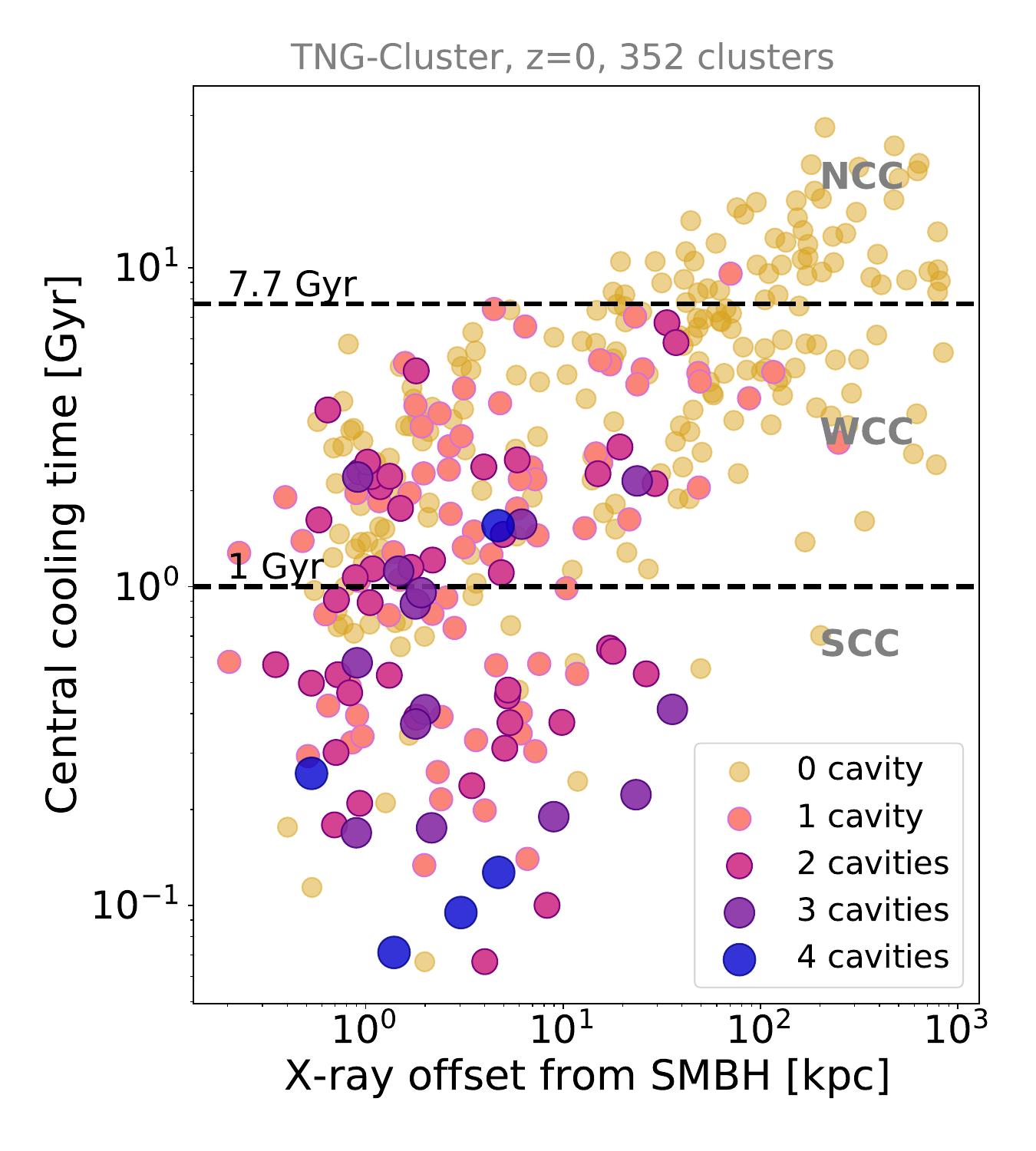}
  \end{subfigure}
  
    \caption{Demographics of clusters with and without identified X-ray cavities in the TNG-Cluster simulation at $z=0$. \textit{Top left:} Percentage of clusters with identified X-ray cavities. \textit{Top right:} Among clusters with identified X-ray cavities, percentage of clusters having one, two, three, or four X-ray cavities. 
    \textit{Bottom Left:} Percentage distribution of cool-core (CC), weak cool-core (WCC), and non-cool-core (NCC) clusters for each category of clusters with or without identified X-ray cavities. The plot shows, for clusters with two X-ray cavities (x-label "2"), that 54 per cent of the TNG-Cluster halos hosting two X-ray cavities are SCCs and 46 per cent WCCs. \textit{ Bottom middle:} Central cooling time vs. X-ray luminosity within $r_{500\text{c}}$ for each halo, color-coded by the number of identified X-ray cavities. \textit{Bottom Right:} Central cooling time vs. the distance between the SMBH and the X-ray luminosity peak, i.e. a measure of un/relaxedness, color-coded by the number of identified X-ray cavities. In TNG-Cluster, $\sim$ 39 per cent of clusters exhibit at least one X-ray cavity. Clusters with identified X-ray cavities tend to have shorter central cooling times and higher X-ray luminosities, and to be more relaxed (smaller X-ray peak offset).}
  \label{fig:panel}
\end{figure*}

\subsection{Demographics}\label{subsec:demogrpahics}

As mentioned above, X-ray cavities are a common phenomenon in TNG-Cluster: here we expand on the frequency and the properties of the simulated clusters hosting X-ray cavities. As shown in the two top panels of Figure~\ref{fig:panel}, according to our visual identification and out of the 352 analyzed clusters, 136 (39 per cent) of $z=0$ TNG-Cluster systems have at least one visible X-ray cavity. In fact, we find a range of simulated clusters containing a variable number of X-ray cavities, from one to four, with the occurrence decreasing as the number of X-ray cavities increases (left panel). More precisely, our study reveals that 52 per cent of the clusters with X-ray cavities have only one identified X-ray cavity, while 29 per cent of the sample have pairs (pie chart).

\textit {Cluster cool-core state}.
Observations suggest that cool-core clusters -- i.e. clusters with short central cooling times --are more likely to host X-ray cavities. \cite{2024Lehle} demonstrated that TNG-Cluster naturally returns a diverse population of cluster cores, from strong cool cores (SCCs, with a central intrinsic cooling time\footnote{for each cluster, a mass-weighted cooling time mean is computed from the simulation output, within a 3D radius of $0.012\times$r$_{500\text{c}}$, and used to determine the CC status.} of the gas below 1 Gyr), to weak cool-cores (WCCs, with central cooling time between 7.7 and 1 Gyr), and non-cool-cores (NCCs), and with relative fractions at $z=0$ in broad agreement with observations. 
Now, according to our analysis, also in TNG-Cluster there is a clear trend between the presence of X-ray cavities and the cool-core nature of the host cluster, as shown in the lower-left histogram of Figure~\ref{fig:panel}. Clusters with identified X-ray cavities are either SCCs or WCCs, and an increasing fraction of SCCs host an increasing number of X-ray cavities in the same cluster. The vast majority of NCCs in TNG-Cluster do not host identifiable X-ray cavities, while over 70 percent of clusters with three or four X-ray cavities are SCCs. Beyond these categories, we find a clear correlation among central cooling time, X-ray luminosity ($L_{\text{X}500\text{c}}$) in the core, and number of X-ray cavities per cluster (bottom middle panel of Figure~\ref{fig:panel}): both at fixed X-ray luminosity in the core and for higher luminosities, multiple X-ray cavities are more frequent in systems with shorter cooling times.

\textit {Cluster relaxedness}.
Observational studies also suggest that X-ray cavities are more frequently associated with relaxed rather than unrelaxed clusters \citep{2023ApJ...954...56O}, whereby the dynamical state of the observed clusters (i.e. whether the system is dynamically relaxed or has recently undergone a merger or interaction with another cluster) is quantified by the offset between the peak of the X-ray emission (indicating the center of mass of the hot gas) and the center of the gravitational potential well. It is not clear whether this relation is the result of causal physical connections between relaxedness and formation or survivability of X-ray cavities, or whether it is a manifestation of an underlying observed correlation between cluster cool coreness and relaxedness. Nevertheless, this link appears to be in place also in TNG-Cluster. In particular, we estimate the relaxedness offset for each of the 352 clusters by comparing the position of the X-ray peak with the location of the SMBH of the BCG, by employing a 'shrinking circle' algorithm to identify the X-ray peak and with larger offsets denoting less relaxed clusters.  
The bottom right-hand panel of Figure~\ref{fig:panel} shows, firstly, that there is a negative correlation between central cooling time and relaxedness in TNG-Cluster systems, i.e. more relaxed clusters tend to have somewhat shorter central cooling times, although with a large scatter. Finally, TNG-Cluster halos with more than two X-ray cavities tend to exhibit smaller (<50 kpc) offsets. More importantly, according to TNG-Cluster, also at fixed cooling time, e.g. in the WCC regime, shorter cooling times seem to be associated with larger X-ray cavity multiplicities, suggesting a connection between X-ray cavity abundance and cluster dynamical state.

\begin{figure*}
\begin{subfigure}[b]{\textwidth}
    \centering
    \includegraphics[width=\textwidth]{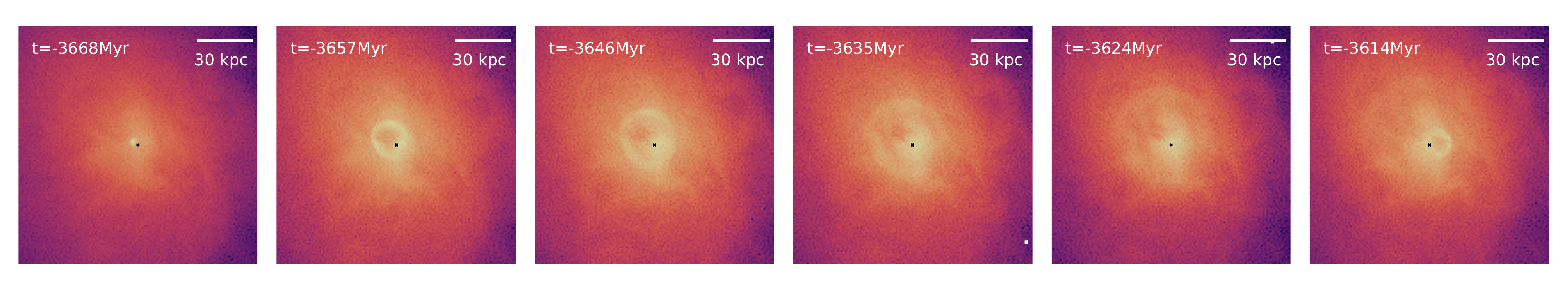}
  \end{subfigure}
  \hfill
  \begin{subfigure}[b]{\textwidth}
    \centering
    \vspace{-0.5cm}
    \includegraphics[width=\textwidth]{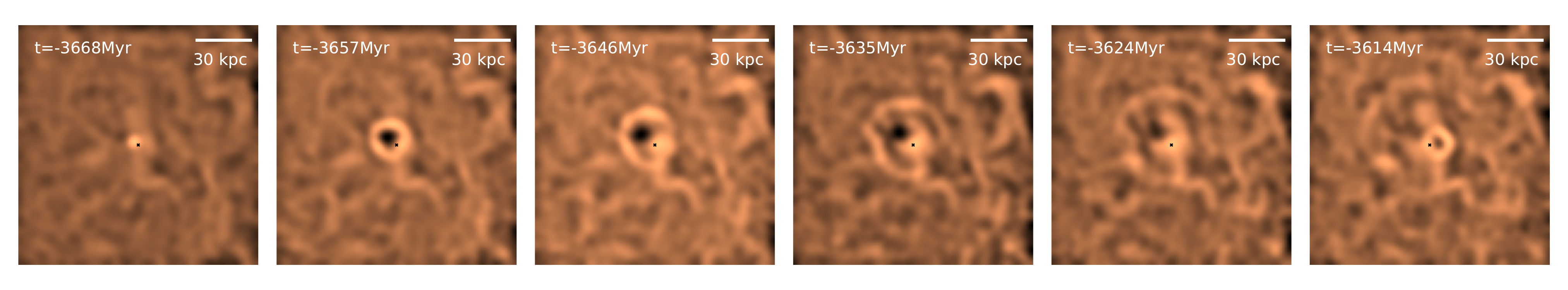}
  \end{subfigure}
    \hfill
  \begin{subfigure}[b]{\textwidth}
    \centering
    \includegraphics[width=\textwidth]{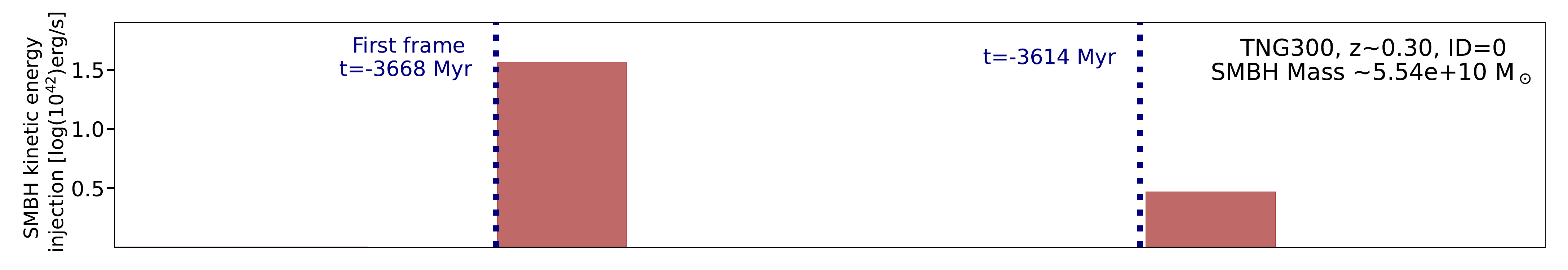}
  \end{subfigure}
      \hfill
  \begin{subfigure}[b]{\textwidth}
    \centering
    \vspace{-0.3cm}
    \includegraphics[width=\textwidth]{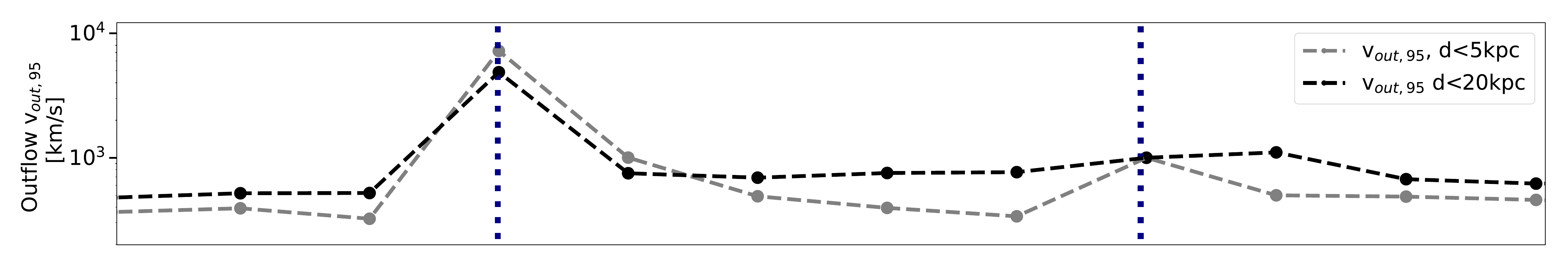}
  \end{subfigure}
        \hfill
  \begin{subfigure}[b]{\textwidth}
    \centering
    \vspace{-0.3cm}
    \includegraphics[width=\textwidth]{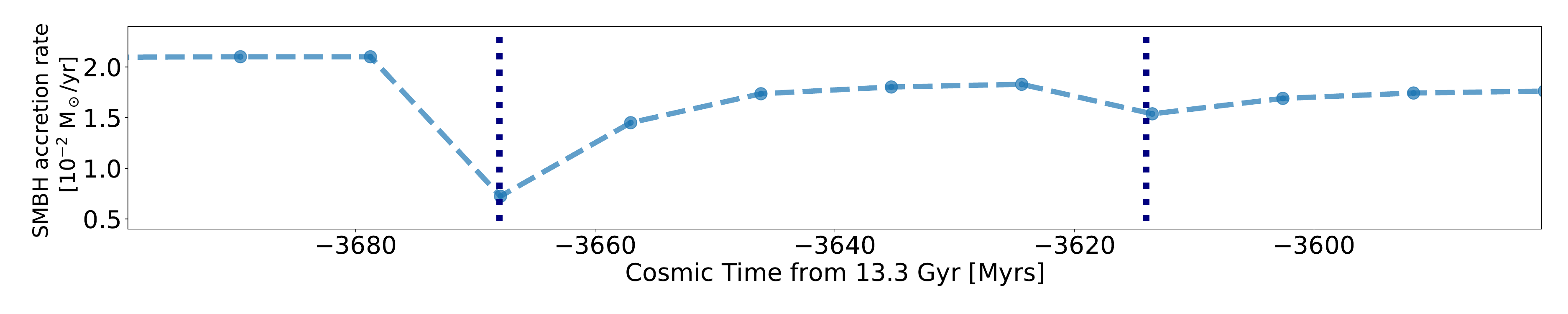
}
  \end{subfigure}
  \caption{Time evolution of selected quantities from the TNG300 most massive cluster, for which high temporal cadence output is available. The top two rows show six X-ray emission maps of the central region of the cluster, across 55 Myr, featuring two X-ray cavity events occurring in the second and last frame: mock Chandra surface brightness maps and the same maps processed with an unsharp mask filter to highlight X-ray-depleted regions, respectively. From top to bottom, the lower three panels quantify the time evolution of: \textit{i)} energy injected in kinetic mode by the SMBH between two successive snapshots, i.e. across time spans of about 10 Myrs; \textit{ii)} maximum radial outflow velocities (95th percentile) of the gas at different distances from the SMBH; and \textit{iii)} SMBH accretion rate. In all panels, the two dotted vertical lines indicate the frames just before the first and second X-ray cavities appear. The two events are characterized by a 10$^{42-43}$ erg s$^{-1}$ kinetic energy release and high-velocity gas outflows, as well as concomitant decreases in the SMBH accretion rate. See continuation in Figure~\ref{fig:subboxe_1gyr}.}
\label{fig:subboxe_panel}
\end{figure*}

\begin{figure*}
    \begin{subfigure}[b]{\textwidth}
    \centering
    \vspace{-0.1cm}
    \includegraphics[width=\textwidth]{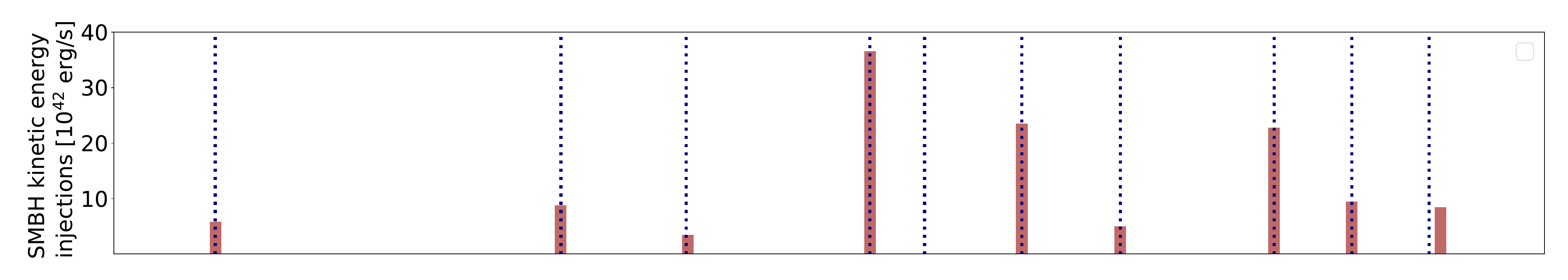}
  \end{subfigure}
      \hfill
  \begin{subfigure}[b]{\textwidth}
    \centering
    \vspace{-0.1cm}
    \includegraphics[width=\textwidth]{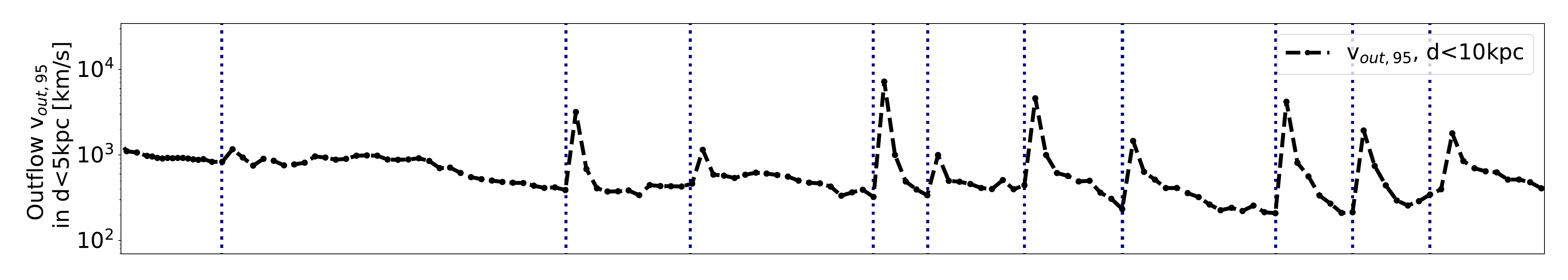}
  \end{subfigure}
        \hfill
  \begin{subfigure}[b]{\textwidth}
    \centering
    \vspace{-0.1cm}
    \includegraphics[width=\textwidth]{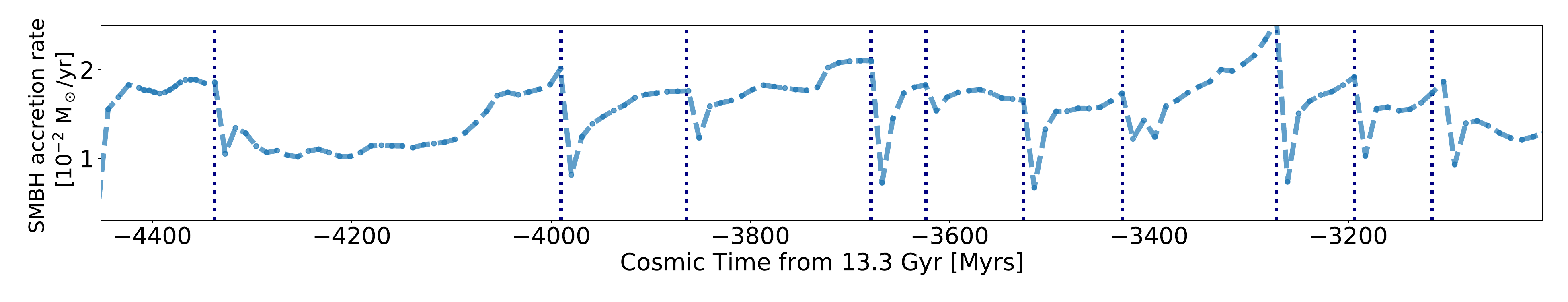}
  \end{subfigure}
  \caption{Time evolution of the same selected quantities as in the previous Figure~\ref{fig:subboxe_panel}, spanning 1.5 Gyr starting from \(z \sim 0.24\). Dotted vertical lines indicate the times when an X-ray cavity is visually identified in the X-ray and UM maps. These X-ray cavities are consistently linked to SMBH energetic outbursts and high-velocity gas outflows. Simultaneously, the SMBH accretion rate decreases in reaction to gas depletion in the SMBH vicinity. These results show the direct connection between the formation of X-ray cavities and SMBH kinetic feedback mode in TNG.}
  \label{fig:subboxe_1gyr}
\end{figure*}

\section{Results: The Origin of X-ray cavities in TNG-Cluster}\label{sec:origin}
So far we have shown that, in the TNG-Cluster cosmological simulation, X-ray cavities are a common manifestation of the underlying galaxy and cluster formation model. In fact, they are also an emergent feature, as no specific TNG model choices or parameters were designed with the explicit intent of reproducing X-ray cavities in the ICM of simulated clusters \citep[see][for all details on the design of the TNG model]{2018MNRAS.473.4077P}. However, we have not yet explored how these structures are created. In the following, we show that within the TNG model, X-ray cavities resembling those observed in actual clusters are formed by kinetic, wind-like energy injections, driven by the SMBHs at the center of clusters. 

\subsection{Insights from the TNG300 cluster subbox}
One way to connect AGN feedback to the creation of an X-ray cavity is to study the behavior of the gas in the vicinity of the SMBH during the X-ray cavity formation period. For this purpose, we use data from the TNG300 simulation \citep{2018bPillepich, 2018Nelson, 2019NelsonPublicReleaseTNG, 2018Springel, 2018Marinacci, 2018Naiman}, in particular its subboxes, which are spatial cutouts of fixed comoving size within the main simulation box and whose data have been stored with high temporal resolution between each subbox snapshots: every $\lesssim$10 Myr, compared to $\sim$150 Myr between the main snapshots. Snapshots at such high time-cadence were not saved for TNG-Cluster, therefore we specifically focus on the TNG300-Subbox-0, which is centered on the most massive cluster of TNG300 (friends-of-friends ID 0), with a mass of \(2 \times 10^{15}\) \msun \hspace{0.05cm} at $z=0$. 
Given that both simulations employ the same TNG galaxy formation and feedback model, have the same resolution, and that this halo mass falls within the range of TNG-Cluster halos (\(M_{500\text{c}} = 10^{14.0}-10^{15.3}\) \msun), it is a representative cluster for our investigation into the physical origin and temporal evolution of X-ray cavities.

In Figure~\ref{fig:subboxe_panel}, we depict two distinct X-ray cavity events formed in this cluster, in a panel of six subbox-snapshots covering a time range of about 55 Myr. We visualize the emergence and progression of these X-ray cavities within 250 by 250 kpc X-ray emission maps (depth of 40 kpc) of the gas centered on the position of the SMBH. We also show the corresponding unsharp-masked maps that enhance the contrast and allow for better visualization. The first panel displays the immediate previous snapshot prior to the first X-ray cavity's appearance. The lower panels of Figure~\ref{fig:subboxe_panel} present the temporal evolution of selected physical quantities from the simulation, where the x-axis represents cosmic time in millions of years from 13.3 billion years after the Big Bang, with time progressing from left to right. In particular, we show the evolution of: \textit{i)} the SMBH kinetic energy injection between snapshots\footnote{As we have not recorded the energy of the individual SMBH feedback injections in the simulation output, here we estimate it by measuring the total kinetic energy gained by the gas cells within a small fixed distance from the center (5 kpc) since the previous snapshot, i.e. over the previous 10 Myrs.}, \textit{ii)} the outflow velocity, measured as the high-end (95th percentile) of the mass-weighted velocity distribution, for the gas cells situated respectively within 5 kpc or 20 kpc from the SMBH, and \textit{iii)} the SMBH instantaneous accretion rate. Additionally, Figure~\ref{fig:subboxe_1gyr} shows the evolution of these quantities over a much longer time period of 1.5 Gyr, to highlight the systematic behaviors associated with the formation of X-ray cavities. By inspecting these time series, we can clearly see temporal correlations among large energy injections from the central SMBH, the development of fast outflows (up to $1000-10000$ km s$^{-1}$), and the emergence of X-ray cavities in the surrounding ICM. We analyze such connection in greater detail in the next section \ref{sec:episodic_bursts_smbh}.

\subsection{X-ray cavities carved by distinct SMBH injection events}\label{sec:episodic_bursts_smbh}

The sequence of images in the upper portions of Figure~\ref{fig:subboxe_panel} shows the formation and development of two individual X-ray cavities corresponding to two instances of kinetic energy released by the central SMBH (\(\Delta E_{\rm kin}\)). A significant increase in \(\Delta E_{\rm kin}\) at the time of X-ray cavity formation is manifest: during the first X-ray cavity event, the kinetic energy release is \(\sim 3 \times 10^{43}\)~erg s$^{-1}$ (between -3668 and -3657 Myr, i.e. between 3668 and 3657 million years ago), while during the second X-ray cavity event (starting at -3624 Myr), such injection is an order of magnitude lower, \(3 \times 10^{42}\) erg s$^{-1}$. Alongside these large kinetic energy releases, there are smaller ones at most times, which however are below the $\sim 10^{41}$ erg s$^{-1}$ level. Additionally, in correspondence to each large kinetic energy injection, there is also a synchronized increase, notably lower ($\sim 10^{38}$ erg s$^{-1}$), in the thermal energy of the gas cells near the SMBH (within d<5kpc, but not shown in the plots), suggesting a partial conversion of kinetic energy to thermal energy.

The second lower panel of Figure~\ref{fig:subboxe_panel} shows that the gas is accelerated and ejected in an episodic manner, achieving outflow velocities of up to several thousands of km s$^{-1}$. After reaching these peak outflow speeds, the outflow velocities gradually decrease over time. The gas at larger distances (from 5 and 20 kpc from the SMBH, for the grey and black curves respectively) exhibits a similar modulation in velocity over time. The first X-ray cavity is linked to synchronized outflow peaks occurring within both 5 kpc and 20 kpc, albeit with a lower maximum value. The second X-ray cavity, formed by a lower-energy kick, exhibits visibly delayed velocity peaks for the gas at progressively larger distances. 
These patterns illustrate the tendency of outflows to decelerate as they travel farther from the SMBH and through the surrounding BCG gas (if any) and the ICM.
Finally, the bottom panel of Figure~\ref{fig:subboxe_panel} shows that X-ray cavity formation is associated with a decrease in the mass accretion rate of the SMBH. The gas expelled immediately after the birth of an X-ray cavity lowers the density near the SMBH,  decreasing the amount of material available for the SMBH to accrete. Therefore $\dot{M}$ is temporarily reduced as gas is expelled from the innermost region of the cluster.

\begin{figure*}
\begin{subfigure}[b]{0.5\textwidth}
    \hspace{-0.2cm}
    \includegraphics[width=\textwidth]{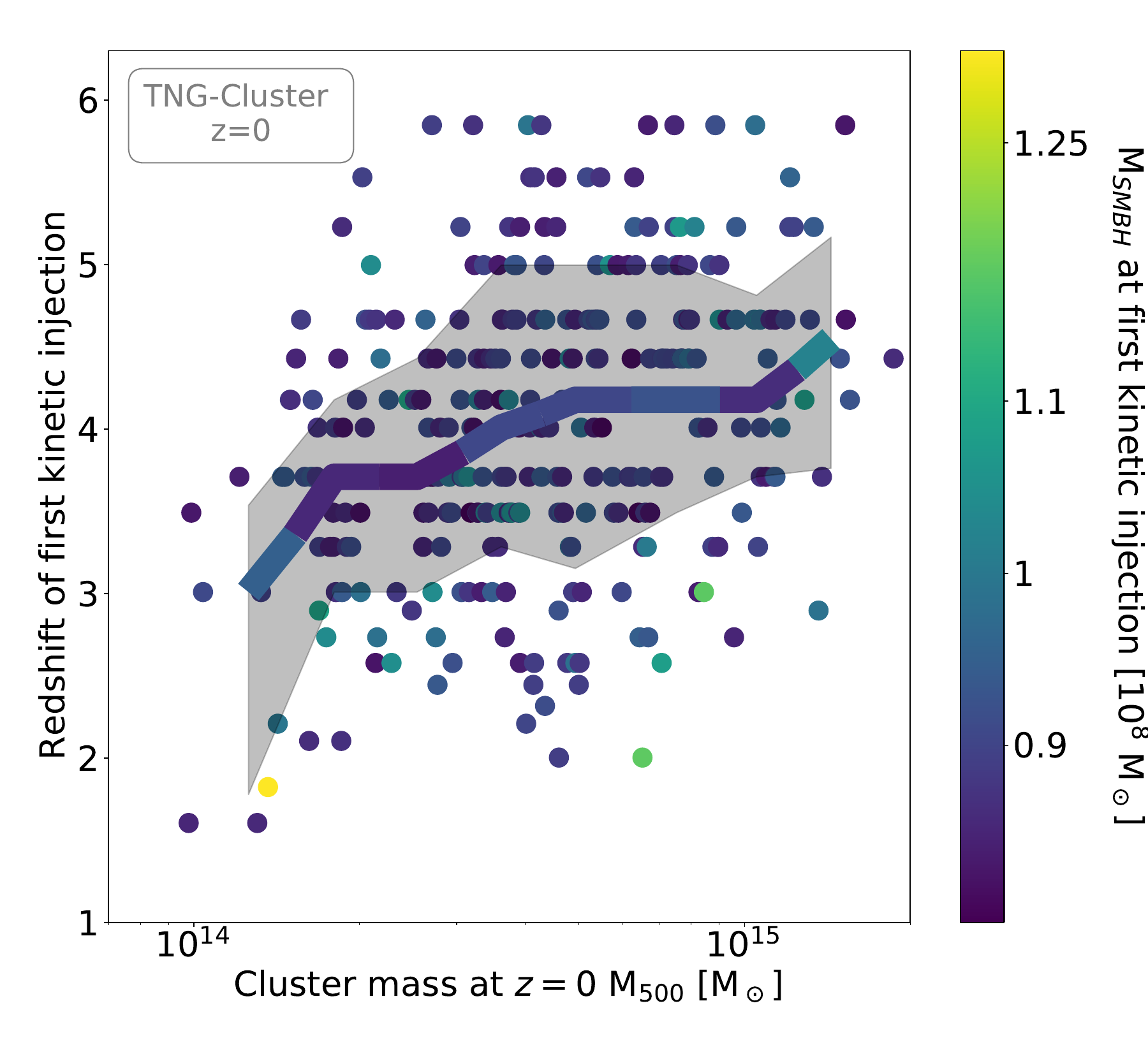}
  \end{subfigure}
  \hfill
  \begin{subfigure}[b]{0.465\textwidth}
    \includegraphics[width=\textwidth]{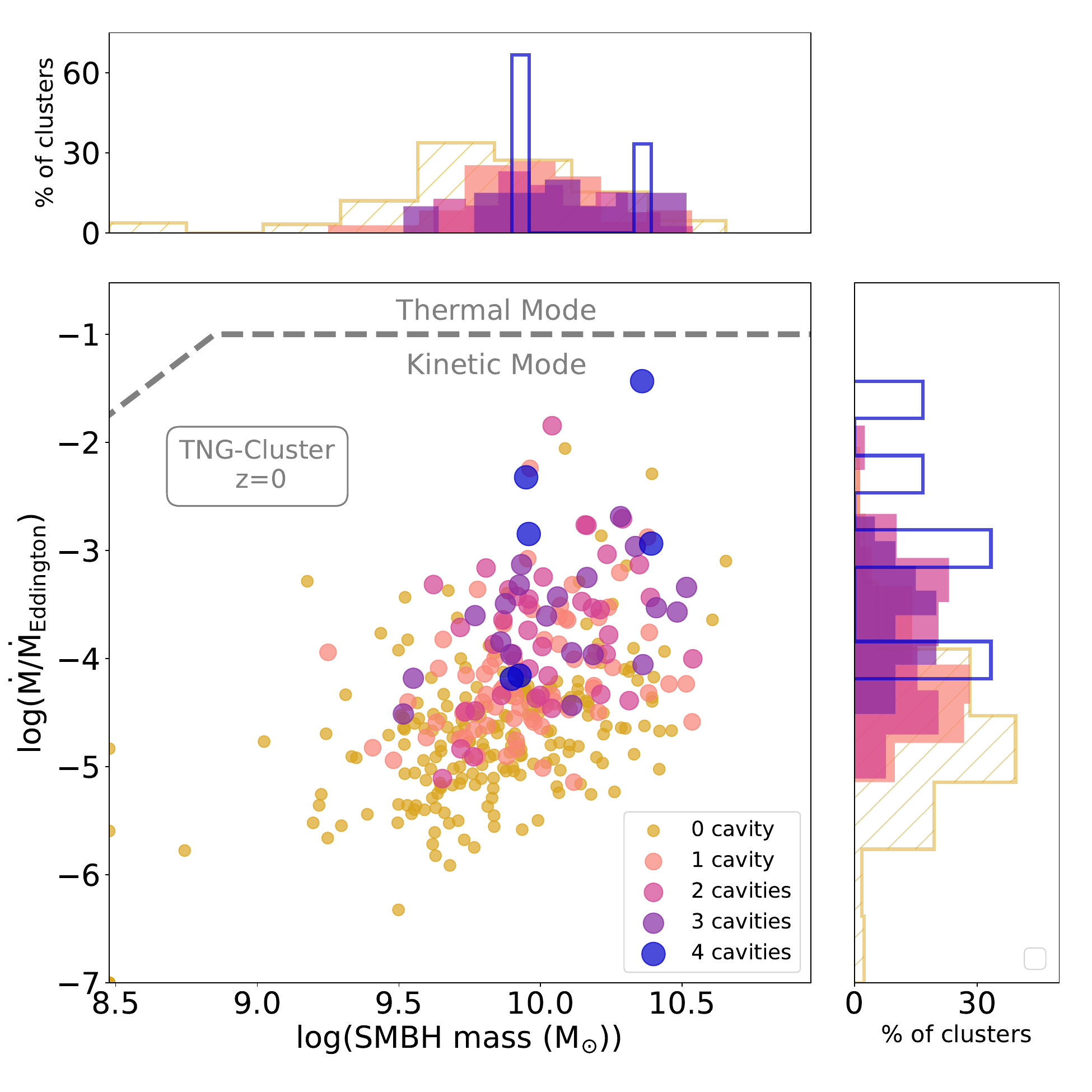}
  \end{subfigure}
  \caption{X-ray cavities as a manifestation of kinetic energy injections from low Eddington-ratios SMBHs according to TNG-Cluster. \textit{Left Panel}: Feedback state of the 352 clusters at $z=0$. The redshift of the first kinetic injection indicates the approximate moment of transition from thermal feedback mode to kinetic mode (curve: median color-coded with the average SMBH mass at the time of transition, grey shaded region: 1-$\sigma$ range of the distribution). All central SMBHs in TNG-Cluster operate in kinetic, low-accretion feedback mode at $z=0$ and since at least $z\simeq1$, with the onset occurring around a SMBH mass of $\sim$ 10$^8$\msun. SMBHs in more massive clusters today ($z=0$) tend to switch to kinetic mode at higher redshifts. \textit{Right Panel}: Fraction of the instantaneous accretion rate relative to the Eddington accretion rate, color-coded with the number of X-ray cavities per cluster. The dashed grey line shows the accretion-mass threshold below which, within the TNG model, SMBHs are in kinetic feedback mode. For visual clarity, data points outside the axis range are displayed at the limits, five SMBHs that fall below these limits overlap in the lower-left corner. Clusters with more massive SMBHs, which accrete at higher rates within the kinetic feedback mode, tend to produce more frequent feedback events, creating multiple co-existing X-ray cavities.}
  \label{fig:fdbckmode}
\end{figure*}

The time evolution of Figure~\ref{fig:subboxe_1gyr} of SMBH kinetic injections, outflow velocities, and SMBH accretion rates on a longer timeline further supports the findings from above: the emergence of X-ray cavities systematically aligns with an SMBH kinetic energy release, synchronized with an outflow velocity peak and a dip in the SMBH mass accretion rate. These results demonstrate the direct connection between the formation of X-ray cavities and the unidirectional momentum kicks given by the SMBH in kinetic feedback mode within the TNG model, at least according to one of the clusters simulated therein. We inspect (but we do not show) the X-ray maps along the evolution of this massive cluster across an even longer period of time, namely 5.5 Gyr (from $z=0.55$ to $z=0$). We choose this range to focus on the last billions of years, close to the current epoch, when the cluster is in a similar state with mass > 10$^{14}$ \msun. Over this period we identify 23 X-ray cavities by visually inspecting the central $250\times250$ kpc region of the halo, giving a mean frequency of one every 240 Myr. The intervals between consecutive SMBH energy injections (or at least those we can capture given the temporal spacing between the subbox-snapshots) range from 40 to 300 Myr. Even if these statistics reflect the evolution of one single object from TNG300, and cannot be a priori generalized, they do suggest that not every kinetic energy injection from the SMBH results in a visible X-ray cavity, as we only capture 23 of them among 35 injection events. We believe that different scenarios for the emergence, properties, and evolution of X-ray cavities, within the same cluster and across different clusters, may depend not only on the amount of injected feedback energy but also on the dynamic state of the gas in the core region, as the maps visually suggest. In a nutshell, the evidence linking the formation of X-ray cavities to the AGN mechanical feedback model in IllustrisTNG can be summarized as follows:

\begin{itemize}
    \item[$\blacksquare$] At the center of the TNG300 most massive cluster we identify numerous X-ray cavities over time, starting from $z\sim$1-2 when the cluster has already switched to the kinetic feedback mode --the SMBH of this system has not exercised thermal mode feedback since $z = 3.5$.
    
    \item[$\blacksquare$] The formation of X-ray cavities along the life of this simulated cluster is systematically associated with an SMBH release of kinetic energy, reaching between $10^{42-45}$ erg s$^{-1}$, and with the emergence of outflows with velocities between 1,000 and 10,000 km s$^{-1}$ (Figure~\ref{fig:subboxe_1gyr}).
    
    \item[$\blacksquare$] The gas expelled immediately after the birth of an X-ray cavity lowers the density near the SMBH, which reduces the mass accretion rate onto the SMBH.
\end{itemize}

Below we argue that this phenomenology, and the causal connection between X-ray cavities and mechanical wind-like feedback, applies throughout our population of simulated clusters.

\begin{figure*}
    \centering
    \includegraphics[width=\textwidth]{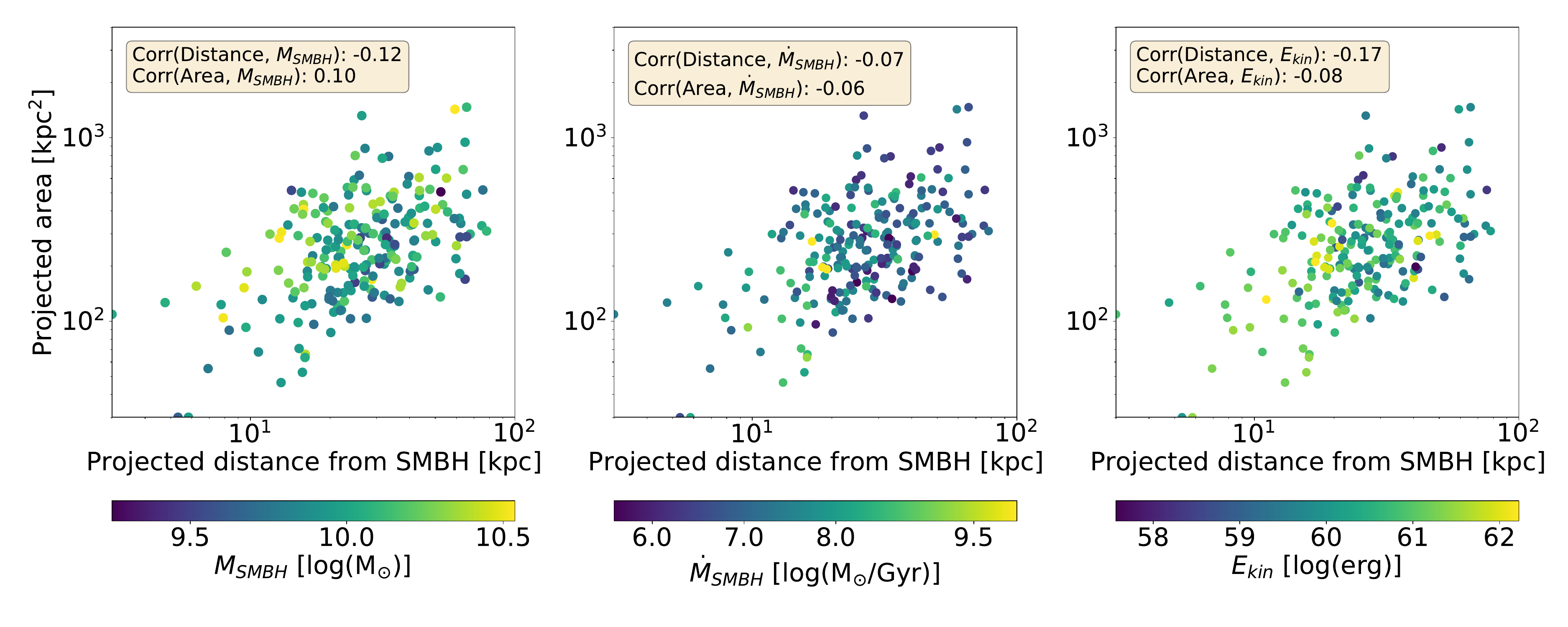}
    \caption{Connections between X-ray cavity area, distance to the SMBH, and various properties of its central SMBH. \textit{From right to left:} points are color-coded with SMBH mass, SMBH instantaneous accretion rate, $E_{\rm kin,z=0}$ amount of kinetic energy injection from the central SMBH at $z=0$. Weak or no correlations can be seen, even though, in TNG-Cluster, X-ray cavities are produced by injections of kinetic energy from SMBHs. This lack of clear correlation highlights that the properties of these structures result not only from SMBH activity but also from a complex interplay with the ICM.}
    \label{fig:cav_area_dist_demogr}
\end{figure*}

\subsection{Link between X-ray cavities and SMBH activity across populations of TNG-Cluster halos}

Focusing again on the whole sample of 352 clusters at $z=0$ from the TNG-Cluster suite, we see in Figure~\ref{fig:fdbckmode} that all their SMBHs are operating in a kinetic, low-accretion feedback mode at $z=0$. In fact, within the TNG model, the SMBHs of very massive systems like those of TNG-Cluster have been at very low Eddington ratios for billions of years and have hence imparted kinetic rather than thermal feedback into the surrounding. The redshift of their first kinetic injection, which approximately albeit not exactly marks the transition from the thermal feedback mode to the kinetic mode, ranges between 5 and 2, with the transition occurring earlier on for more massive clusters (left panel of Figure~\ref{fig:fdbckmode}). We have checked and confirmed that the SMBHs of the $z=0$ BCGs have been exclusively in kinetic mode since at least $z=1$ (with rare exceptions): the TNG-Cluster X-ray cavities at $z=0$ studied and presented in this paper are therefore phenomena that emerge in the presence of kinetic unidirectional energy injections from the SMBHs. 

In the right panel of Figure~\ref{fig:fdbckmode} we inspect the ratio of the SMBH mass accretion rate to the Eddington rate ($\dot{M}_{\rm SMBH}/\dot{M}_{\rm Edd}$, where $\dot{M}_{\rm Edd} = \frac{L_{\rm Edd}}{{\epsilon_r}}$ with $\epsilon_r = 0.1$ the radiative efficiency parameter) plotted against the mass of the BCG's SMBH of each cluster. The markers are color-coded based on the number of X-ray cavities identified in these clusters, ranging from zero to four. The mass accretion rate depicted in the figure is instantaneous, capturing the state of SMBHs at a snapshot in time, despite some X-ray cavities might have been inflated tens or even hundreds of Myr earlier. All clusters at $z=0$ are undergoing kinetic feedback mode and typically host SMBHs that accrete at low rates (10$^{-6}$ to 10$^{-2}$ of the Eddington accretion rate), with more massive SMBHs exhibiting higher accretion rates. The figure reveals a trend whereby clusters with more numerous simultaneous X-ray cavities show higher accretion rates and SMBH masses. These findings suggest that SMBHs accreting at higher rates(while remaining within the low-luminosity i.e. kinetic feedback regime), and therefore growing faster, may generate more frequent successive feedback events carving multiple X-ray cavities. 

In Figure~\ref{fig:cav_area_dist_demogr}, we further explore how SMBH mass, accretion rate, and amount of kinetic energy injected at the current $z=0$ snapshot $E_{\rm kin,z=0}$ may correlate with the size and distance of the X-ray cavities of TNG-Cluster. However, there does not seem to be a clear correlation between the properties of SMBHs and those of the X-ray cavities. It is important here to note the temporal disparity: some distant X-ray cavities might have formed several tens of Myr before the recording of the SMBH's accretion state or $E_{\rm kin,z=0}$. Furthermore, the variability in timing between subsequent kinetic energy injections can range from 0.7 Myr up to a few hundred Myr for the same SMBH. This temporal variability and disparity add another layer of complexity when trying to correlate the characteristics of X-ray cavities (which are expected to be influenced by these energy injections) with the current activity and mass of the SMBH. Additionally, the area and ascent velocities of the X-ray cavities are probably influenced by a complex interplay of ICM density, pressure, and motions, on top of the SMBH activity. The lack of correlation and interpretability of Figure~\ref{fig:cav_area_dist_demogr} underscores the complexity of these connections, even in the case of the simulations where the causal physical link is clear(er). Connecting the observed characteristics of X-ray cavities to the current properties of SMBHs might not be as informative after all, or any straightforward interpretation should be put forward with caution.
In the TNG model, not only the most massive BCGs but also massive galaxies in lower-mass clusters and groups experience mechanical feedback from their SMBH, which serves as the dominant and effective energy source within these galaxies \citep{2017RainerW, 2018Weinberger}. This feedback is, in fact, responsible for quenching star formation \citep[e.g.][]{2018Nelson, 2020Terrazas, 2020Zinger, 2020Davies} in the vast majority of these galaxies, including those in TNG-Cluster \citep{2024Nelson}. In terms of feedback energy injections, this overall picture leaves no other plausible explanation than SMBH kinetic feedback for the emergence of the X-ray cavities revealed in this paper. 
Overall, these results are in agreement with the study of \cite{2021PillepichErosita} on eROSITA-like bubbles in TNG50 Milky Way-like galaxies, which suggests that these features result for at least the vast majority of the systems, from kinetic energy injections from SMBHs at the center of galaxies. The X-ray cavities in the ICM of TNG clusters are formed by similar processes, i.e. by intermittent, unidirectional, powerful energy outbursts from the central SMBH. The synchronization between the energy released from the SMBHs into the gas and the peaks of high-velocity outflows seen in the previous Sections highlights the episodic nature of the TNG SMBH feedback and its significant impact on the surrounding gas dynamics. 
Whereas the formation of X-ray cavities is triggered by gas outflows in turn due to SMBH kinetic kicks, their evolution in time is likely influenced by a combination of factors, including SMBH activity and the turbulent, complex physical state of the ICM at the cluster centers \citep[][]{2023Ayromlou}. Because of this richness of phenomenology, it may actually be very hard to pin down, e.g. in observations, a causal link between SMBHs and X-ray cavities.

\section{Discussion} \label{sec:discussion}

How do the overall occurrence and properties of TNG-Cluster X-ray cavities compare to those found in observations? If SMBH feedback determines the formation of X-ray cavities in the TNG model, what may be the impact of the ICM environment on their subsequent evolution and survival? And finally, are X-ray cavities the only heating channel of the ICM in TNG-Cluster? 

\subsection{Thoughts on simulated vs. observed X-ray cavities}\label{sec:discussion_comp}
In this section, we comment on the possible similarities and differences between simulated X-ray cavities in the TNG-Cluster cosmological framework and those observed in real clusters. However, this discussion is merely qualitative -- a detailed, in-depth, and quantitative comparison is given in the second article of this series (Prunier et al. in prep), where we systematically account for selection effects and instrument-related biases.

\paragraph*{Frequency i.e. detection rate.} In the TNG-Cluster simulation, 39 per cent (136/352) of the BCGs exhibit X-ray cavities, which is comparable to the 52 percent (69/133) reported by the observational study of \cite{2016Shin} and 43 percent (15/35) by \cite{2014Panagoulia2}, both in systematic searches across cool-core and non-cool-core clusters. However, this frequency is lower than that of other studies which have reported higher detection rates \citep[e.g.,][]{2006DunnFabian,2006Rafferty,2012Birzan}, although those studies are often biased toward the brightest sources in the sky. It should be noted that the detection rate in our systematic study of each TNG-Cluster system could be influenced by several factors, including many that are unrelated to the TNG SMBH feedback prescription. Unlike most observational studies, which typically have exposure times of 30-50 ks, our study benefits, by design and construction, from longer exposure times of 200 ks. This extended duration, coupled with the placement of clusters at a consistent angular distance of 200 Mpc from the detector, enhances our ability to identify X-ray cavities. The longer exposure reduces the likelihood of missing X-ray cavities that might be overlooked in shorter exposures and mitigates redshift-angular resolution effects. Furthermore, during the detection process on our mock Chandra images, it was sometimes challenging to differentiate between X-ray cavities and other X-ray surface brightness depressions caused by gas sloshing. This effect could affect the number of detected X-ray cavities, though this issue also arises in real observations.

\paragraph*{Pairs.} In TNG-Cluster we identify halos hosting one, two, three, and up to four simultaneous X-ray cavities, although the occurrence of larger numbers becomes increasingly rare. In observations, X-ray cavities in clusters are also detected either alone \citep[e.g., Abell 1795,][]{2014MNRAS.445.3444W}, in pairs \citep[e.g., Abell 2597,][]{2001ApJ...562L.149M}, or in larger numbers such as in Hydra A, \citep[][]{2007ApJ...659.1153W} and M87 \citep[][]{2016Shin} with six X-ray cavities each. One notable difference between the TNG-Cluster findings and observational studies lies in the frequency and spatial configuration of X-ray cavity pairs. Our analysis reveals that 52 per cent of simulated clusters with X-ray cavities host only one identified X-ray cavity, while 29 per of our TNG-Cluster sample exhibit pairs (cf pie chart in Figure~\ref{fig:panel}). This distribution suggests that single X-ray cavities are more prevalent in TNG-Cluster compared to clusters with two or more. In contrast, \cite{2016Shin} found 55 per cent of X-ray cavity pairs and 26 per cent of single cluster, \cite{2004Birzan} detected pairs in 83 per cent of their sample, and \cite{2010ApJ...712..883D} found 50 per cent. Moreover, according to these studies, X-ray cavity pairs in observations are typically symmetrically positioned relative to the AGN center, as they usually are associated with detected bipolar AGN jets. In our TNG-Cluster sample, only 13 pairs of X-ray cavities (11 per cent) exhibit a marked axi-symmetry with respect to the central SMBH (e.g., TNG-Cluster ID 15500142 in Figure~\ref{fig:panel_cav}). As discussed in Section \ref{sec:episodic_bursts_smbh}, TNG X-ray cavities arise naturally from the SMBH kinetic feedback model. The scarcity of X-ray cavity pairs and symmetry in our study likely stems from the TNG model's prescription for AGN feedback, where energy injection occurs in a unidirectional manner. Whereas it is notable that somewhat symmetric X-ray cavity pairs do in fact emerge from independent energy injection events \citep[see also a related discussion in the case of eROSITA-like pairs in Milky Way-like galaxies in the TNG50 simulation,][]{2021PillepichErosita}, the nature of the TNG SMBH feedback, and the shortage of e.g. dense gaseous disks capable of re-directing the gaseous outflows in preferential directions around the typical BCG's SMBH, naturally leads to a preferential formation of asymmetric and unique X-ray cavities.

\paragraph*{Shapes and sizes.} In TNG-Cluster, we identify X-ray cavities at various evolutionary stages, from inflating to rising bubbles, using visual inspection and flagging to distinguish their configurations and morphologies. Although real X-ray cavities are often approximated as ellipsoids, studies such as \cite{2006Rafferty,2012Hlavacek} reveal that observed X-ray cavities have more complex structures than simple ellipses. They tend to be elongated either along the jet direction or in the perpendicular direction, a behavior supported by idealized hydrodynamical simulations of mechanical AGN feedback \citep{2009arXiv0905.4726B,2012ApJ...750..166M,2015ApJ...803...48G}. TNG-Cluster X-ray cavities are more regular and spherical when small and attached to the central SMBH. Larger, more distant X-ray cavities are often deformed in arc-like shapes or elongated toward the SMBH. The flagged "larg" X-ray cavities at the bottom of Fig.~\ref{fig:panel_cav}, surrounding the SMBH, are likely being inflated along our line of sight, suggesting that the energy injection is directed towards us. Some of them exhibit jellyfish-like shapes, like the rising X-ray cavity in TNG-Cluster ID 16921354 of Figure~\ref{fig:panel_cav_phys_pro}, analogous to observed mushroom-shaped X-ray cavities such as in M87 \citep{Churazov2001}.
This variety suggests that the dynamics and evolution of X-ray cavities in the TNG-Cluster simulation share similarities with observational findings, where X-ray cavity shapes are influenced by the feedback processes as well as by their interaction with the surrounding ICM (see also Section \ref{sec:discussion_ICM}).

When exploring the sizes of X-ray cavities in TNG-Cluster, we find that the distribution of mean radii spans from a few kiloparsecs to a few tens of kiloparsecs, which also aligns with the size ranges reported in observational studies \citep[e.g.,][]{2014Panagoulia2}. Our analysis of the connection between the projected distance from the central SMBH and the X-ray cavity area, as shown in Figure~\ref{fig:area_distance}, reveals a positive linear correlation. This indicates that as the distance from the SMBH increases, the area of X-ray cavities tends to grow. While our findings suggest a moderate correlation (r$_p$<0.5), several observational studies, including those by \cite{2016Shin}, \cite{2004Birzan}, and \cite{2008diehl}, have reported a stronger positive correlation between X-ray cavity area (or size) and distance.

\paragraph*{Comparison to other cosmological simulations}
When juxtaposing the projected X-ray cavity area versus distance contours from the Hyenas simulations \citep{2024Jennings} with TNG-Cluster (Fig.~\ref{fig:area_distance}), we find that both cavity populations exhibit similar size ranges, though with different slopes and some larger >$10^3$kpc$^2$ cavity areas in Hyenas and RomulusC. Notably, the Hyenas simulation suite focuses on X-ray cavities within galaxy groups, not clusters. However, their areas and distances are larger and further than those typically found in observed groups, and align more closely with the X-ray population observed in clusters. Hyenas uses the SIMBA model, in which kinetic feedback is implemented in a bipolar fashion, with jet axes aligned with the angular momentum of the inner gas disk. In contrast, the RomulusC simulation \citep{2019Tremmel} does not have yet a detailed characterization of its X-ray cavities, but such cavities appear in the X-ray maps with morphologies resembling those seen in real clusters. RomulusC uses a purely thermal feedback model with continuous energy transfer and shut-off cooling, rather than kinetic feedback. These results suggest that different AGN feedback models can produce X-ray cavities, supporting the realism of the simulation predictions at small scales. However, precisely because different types of energy injections from the SMBHs can result in X-ray cavities, all this indicates that their mere presence is not a strict constraint on the models' choices. A more quantitative comparison between simulated and observational X-ray cavities is needed for deeper insights.

\paragraph*{Thermodynamics of the gas.} 
Figure~\ref{fig:panel_cav_phys_pro} showcases the typical gaseous properties within and around identified X-ray cavities. Throughout the TNG-Cluster sample, we identify recurring patterns, including their manifestation as lower-emitting X-ray regions filled with hot gas (10$^{7.8-8.1}$K) and lower density. As quoted above, about 25 per cent of the X-ray cavities in TNG-Cluster exhibit bright and overpressurized dense edges, coinciding with weak shocks, while X-ray cavities lacking bright edges or detached from the central SMBH show no evidence of shocks. It is important to note that these considerations are based on maps such as the one presented in Fig.~\ref{fig:panel_cav_phys_pro}, which are derived from quantities outputted by the simulation rather than inferred from the mock Chandra observations. On the other hand, in observations, gas properties such as temperature or pressure at X-ray cavity sites are measured by spectral fitting and are limited by instrumental constraints. Consequently, drawing comparisons between these simulation outputs and actual observational results must be made with caution. Keeping this in mind, we can tentatively say that the gas properties of TNG-Cluster's X-ray cavities have similarities to those expected for real ones. They appear to be similarly underdense and exhibit a comparable range of X-ray brightness depletions with respect to the ICM (30-40 per cent e.g., \cite{2015Hlv,2016Shin}. The absence of strong shocks at their boundaries also aligns with observations. Observational studies suggest that buoyancy, rather than excess pressure, drives their outward expansion. Similarly, in TNG-Cluster, X-ray cavities seem to be in pressure equilibrium and rise and increase in volume as they detach from the central SMBH. Observationally, it remains unclear whether X-ray cavities are filled with hot gas, as seen in the TNG simulations: observationally, the high contrast between the density of the X-ray cavities and that of the ICM has been used to constrain the temperature of the thermal plasma, potentially supporting a picture with X-ray cavities gas temperatures of $\gtrsim 2-5 \times 10^8$ K \citep{2002Nulsen,2002MNRAS.337...71S,2007Sanders}.

\paragraph*{Correlation with cool-core status.}
Observational studies have shown that, in CC clusters, rapid cooling at the center is balanced by AGN heating. In this scenario, strong central cooling of the ICM feeds the SMBH in the central galaxy and triggers feedback that creates energetic X-ray cavities. The energy released by these X-ray cavities is believed to offset the cooling process \citep[e.g.,][]{2006MNRAS.366..417F} and aligns with the observed rate of star formation. In TNG-Cluster, a clear trend exists between the presence of X-ray cavities and the cool-coreness of the host cluster, as depicted in the lower-left histogram of Figure~\ref{fig:panel}: simulated clusters with identified X-ray cavities are either SCC or WCC, and clusters with shorter cooling timescales are brighter in X-rays and harbor more X-ray cavities. In the observational literature, X-ray cavities are most frequently found in CC clusters, with approximately two-thirds of CCs displaying clear X-ray cavities, as noted by \cite{2005Dunn}. In contrast, NCCs generally do not exhibit identifiable X-ray cavities, as reported by \cite{2010ApJ...712..883D}. Selection effects might influence the observed prevalence of X-ray cavities in CC clusters: clusters with shorter cooling times are brighter in X-rays, making the X-ray cavities easier to identify due to the higher contrast with the surrounding ICM \citep[CC bias][]{2011A&A...526A..79E}. This selection effect also affects our study of TNG-Cluster X-ray cavities as it is based on observer-like methods on Chandra mocks. Nevertheless, the high rate of identified X-ray cavities in strong CC TNG-Cluster halos supports the idea of a self-sustaining feedback loop scenario.

\paragraph*{Impact of cluster dynamical state.}
Some observational studies suggest that the dynamical state of a cluster, describing whether it is relaxed or has recently undergone a minor-major merger or interaction with another cluster, may influence the presence and identification of X-ray cavities \citep[e.g.,][]{2023ApJ...954...56O}. Relaxed clusters, characterized by stable conditions, typically exhibit persistent activity of central SMBHs, which create and maintain X-ray cavities through energetic outbursts. In contrast, X-ray cavities in unrelaxed clusters are expected to be more prone to disruption by turbulent motions or mergers. Furthermore, the consistent X-ray background in relaxed clusters makes X-ray cavities easier to identify, as the depressions in X-ray emission stand out more clearly. In the TNG-Cluster simulation, relaxed clusters tend to host more X-ray cavities, suggesting that: \textit{i)} the disturbed ICM may impact the longevity of these X-ray cavities, as fewer are identified in unrelaxed clusters, and \textit{ii)} multiple X-ray cavities are more likely to result from successive SMBH outbursts rather than ICM disturbances fragmenting existing X-ray cavities. 

\paragraph*{Misalignments.} 
In observations of real clusters, pairs of older and younger X-ray cavities are often aligned along a common axis, suggesting that the jets from the central AGN have maintained a consistent direction over time \citep[e.g.,][]{ 2007ApJ...659.1153W,2015ApJ...805..112R}. However, misalignments are also observed between the central AGN jets and X-ray cavities \citep{2024Ubertosi}. Several scenarios have been investigated to explain off-axis X-ray cavities: jet realignment over time \citep[e.g.,][]{2021Schellenberger,2021Ubertosi}, strong magnetic fields bending or deflecting the jets and interactions between infalling cold gas and the outflowing AGN jets \citep[e.g.,][]{2021Chibueze_bent_jet,2024Fournier}, but also ICM motions distorting the dynamics of jets, leading to the creation of misaligned X-ray cavities \citep[e.g.,][]{2012Mendygral}. In the TNG simulations, both the random directions of the SMBH kinetic injections and the turbulent movement of gas near the X-ray cavities can influence their position. Additionally, we speculate that sloshing spiral patterns in the central region of the BCG significantly impact the trajectory of the rising X-ray cavities, causing them to follow a spiral path and deviate from their original inflation direction. We expand below on the impact of the surrounding ICM onto the X-ray cavity properties and evolution.

\begin{figure*}
\begin{subfigure}[b]{\textwidth}
    \centering
    \includegraphics[width=\textwidth]{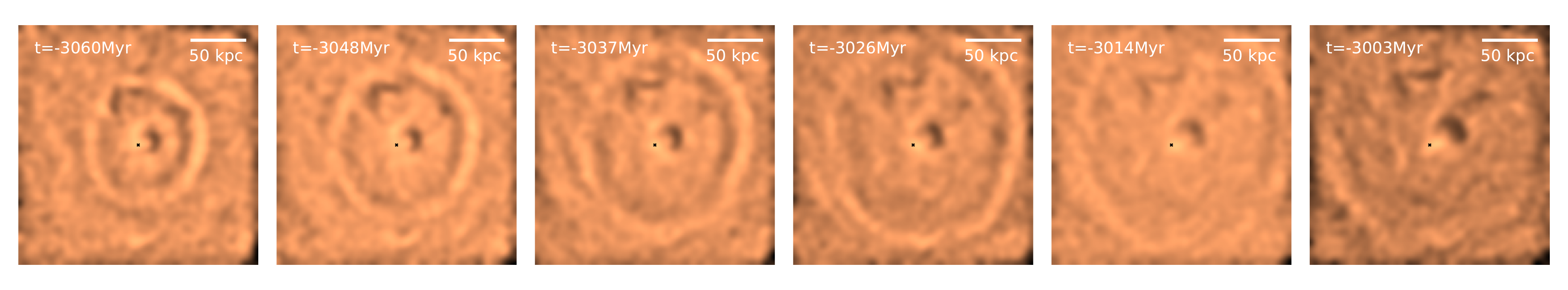}
  \end{subfigure}
      \hfill
  \begin{subfigure}[b]{\textwidth}
    \centering
    \vspace{-0.4cm}
    \includegraphics[width=\textwidth]{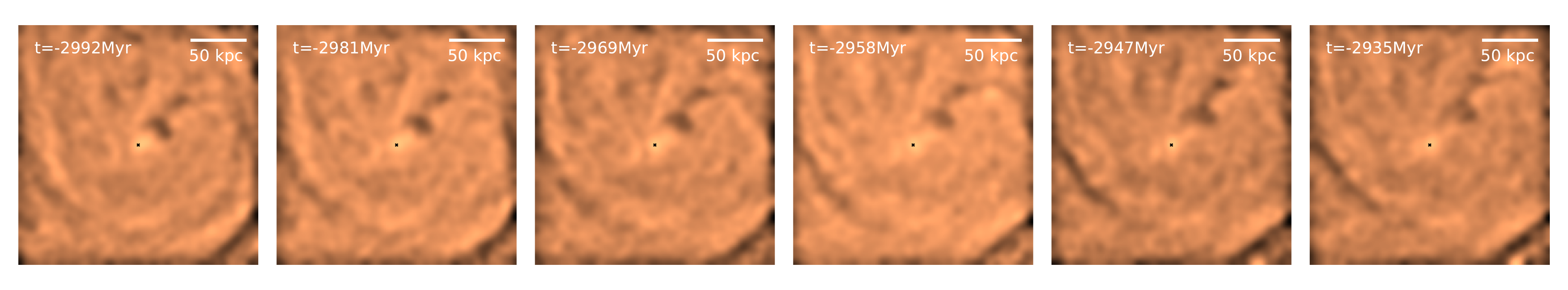}
  \end{subfigure}
        \hfill
  \begin{subfigure}[b]{\textwidth}
  \vspace{-0.4cm}
    \centering
    \includegraphics[width=\textwidth]{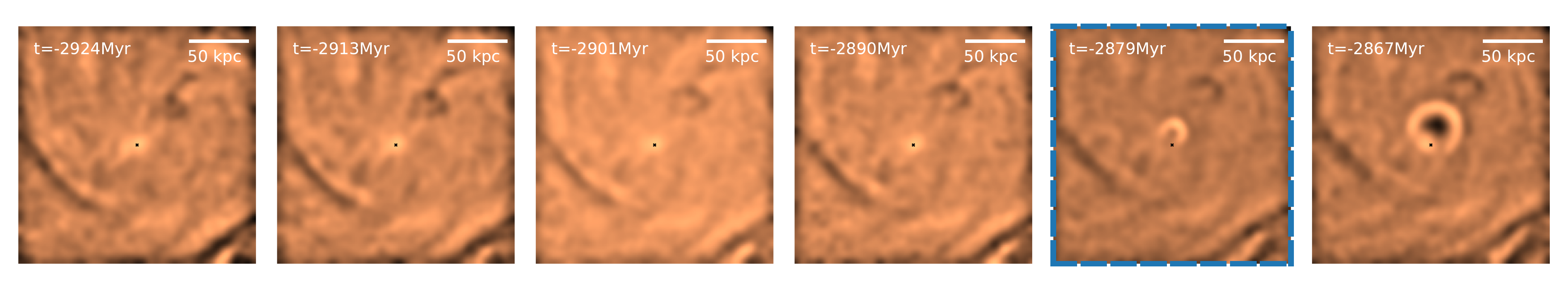}
  \end{subfigure}
    \hfill
  \begin{subfigure}[b]{\textwidth}
    \centering
    \vspace{-0.4cm}
    \includegraphics[width=\textwidth]{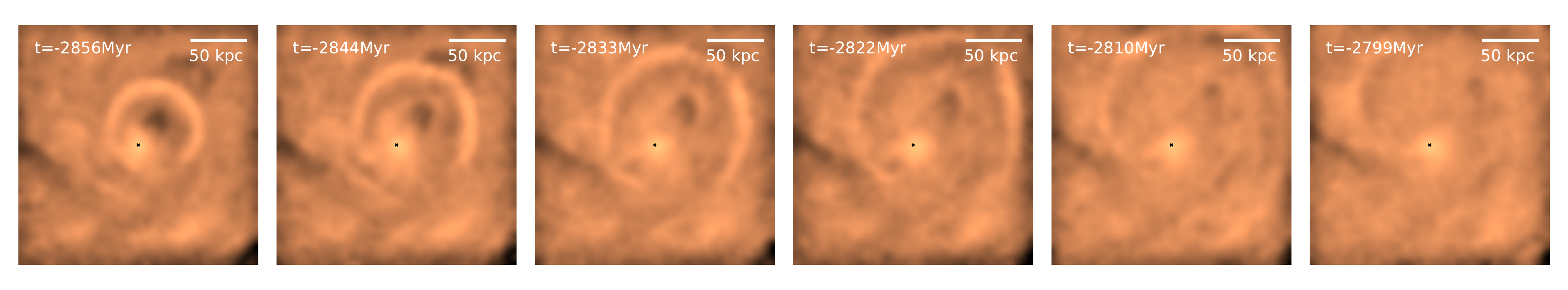}
  \end{subfigure}
    \begin{subfigure}[b]{0.32\textwidth}
    \includegraphics[width=\textwidth]{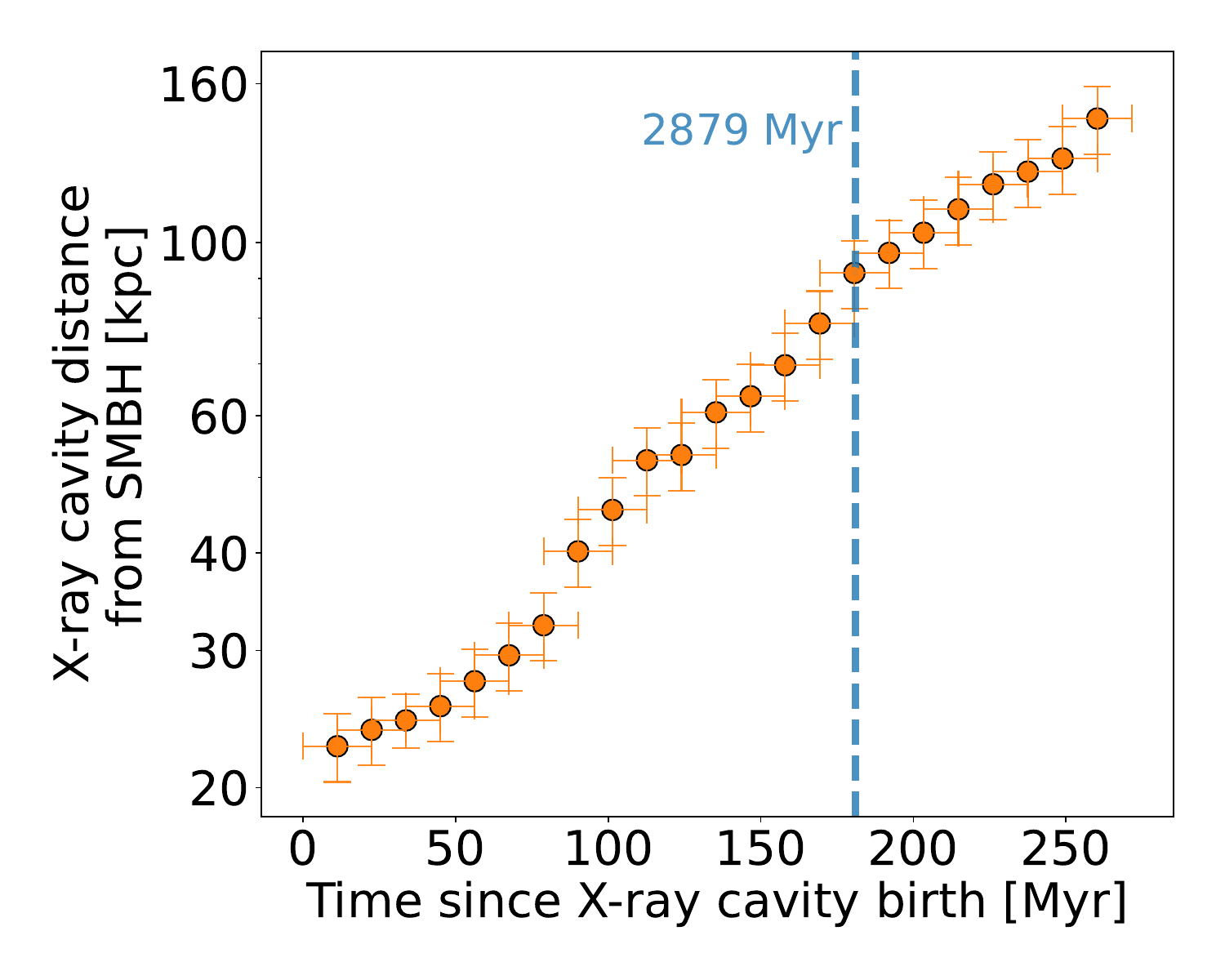}
  \end{subfigure}
  \hfill
  \begin{subfigure}[b]{0.32\textwidth}
    \includegraphics[width=\textwidth]{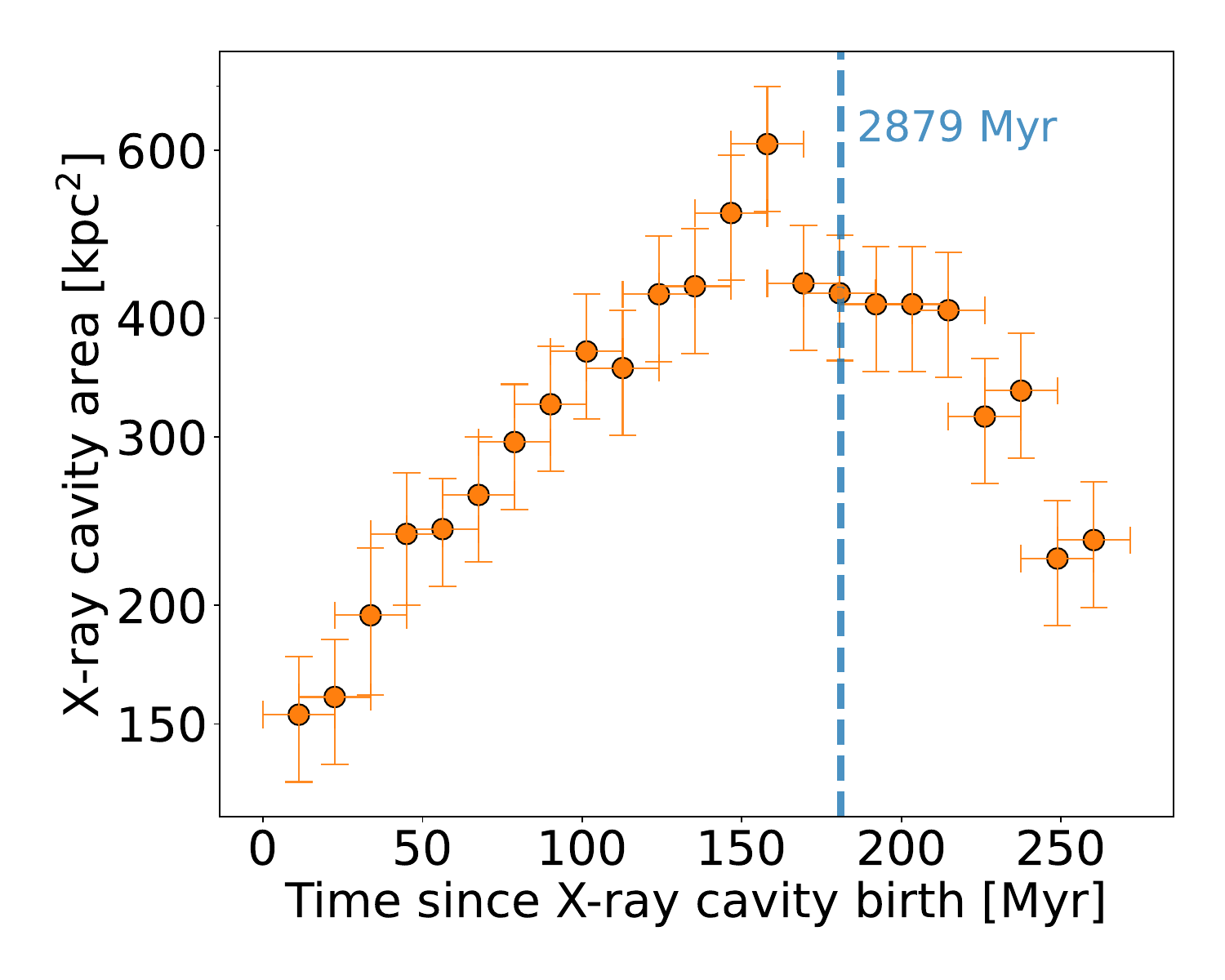}
  \end{subfigure}
    \hfill
  \begin{subfigure}[b]{0.32\textwidth}
    \includegraphics[width=\textwidth]{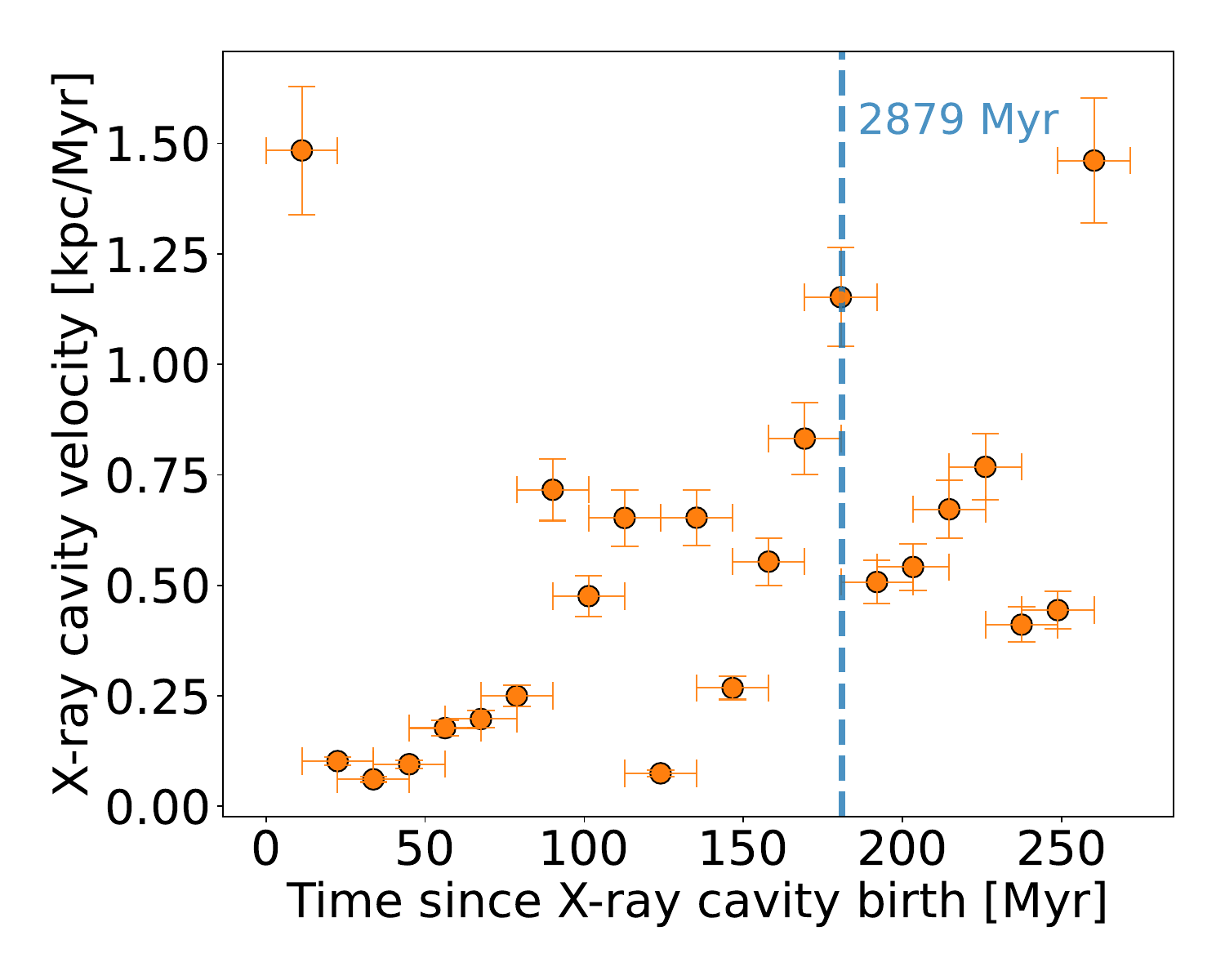}
  \end{subfigure}
  \caption{Example of an X-ray cavity life cycle in the most massive halo of TNG300: from birth till disruption by another injection event. The time difference between each snapshot is $\sim$10 Myr. \textit{Bottom left:} Projected distance from the SMBH. \textit{Bottom middle:} Evolution of the projected area. \textit{Bottom right:} Mean velocity in each frame. We can witness the birth of one X-ray cavity at -3060 Myr from today, the X-ray cavity detaches progressively from the SMBH and rises radially behind the pressure wave of the first kinetic injection. At t= -2879 Myr, a second energy injection gives rise to another X-ray cavity, while the first one is disrupted by the associated front at t= -2822 Myr. This scenario is typical in a calm ICM.}
  \label{fig:subboxe_2063}
\end{figure*}

\subsection{Time evolution of TNG-Cluster X-ray cavities and influence of the ICM weather}\label{sec:discussion_ICM}

Thanks to the cosmological simulation setup of TNG-Cluster, we can argue that TNG-Cluster X-ray cavities are sensitive to the ICM ``weather'', where by weather we refer to dynamic processes within the hot intracluster gas. These include turbulence, sound waves, and weak shocks from galaxy motions and AGN activity but also cold fronts and sloshing motions from mergers. 

The ICM can exhibit various conditions in the cluster core, such as calm and undisturbed phases, which allow X-ray cavities to rise buoyantly and persist for extended periods (up to a few hundred Myr). Conversely, during turbulent phases, the ICM can disrupt or erase X-ray cavities shortly after their formation. This behavior is illustrated in Figure~\ref{fig:subboxe_2063}, where we track the spatial evolution of a long-lived X-ray cavity in the TNG300 most massive halo subbox (top panel) along with its corresponding area, distance from the SMBH, and rising velocity time evolution (bottom three panels). X-ray cavities tend to rise and grow in size over time with various slopes indicative of the X-ray cavities' ascent dynamics. A steep slope indicates a rapid, radial ascent through relatively undisturbed ICM, allowing the X-ray cavity to travel farther. Conversely, a gentler slope suggests a slower-rising X-ray cavity. However, we also identify many examples where the X-ray cavity's path is influenced by surrounding gas motion, resulting in a less radial ascent (Figure~\ref{fig:panel_cav}, TNG-Cluster ID 4753637 and 17352215). However, in the final stages, their areas visually decrease or stop growing as they become fainter and start blending with the ICM's background emission, or as they are being erased by a sound wave or weak shock front, like the X-ray cavity of Figure~\ref{fig:subboxe_2063} at -2810 Myr. These results reflect the dynamic and complex interactions between the AGN-inflated X-ray cavities and the intracluster gas and the timing of AGN feedback events in the simulation. 

Here are the principal scenarios that we have identified by inspecting the TNG300-subbox output in terms of X-ray cavities behavior in the ICM after they detached:

\begin{itemize}
    \item \textbf{Calm ICM}: In periods where the ICM seems visually undisturbed, X-ray cavities have the time to rise radially. We can observe them for several tens of Myr (>50 Myr up to a few hundreds of Myr) after their inflation. As these X-ray cavities rise, they seem to fade away gradually.
    \item \textbf{Disturbed ICM with Sloshing}: During periods of sloshing, where spiral patterns are present in the ICM, X-ray cavities tend to follow these spiral paths and fade quickly.
    \item \textbf{Minor Mergers or Strong ICM Disturbances}: During minor mergers or significant disturbances in the ICM, X-ray cavities are often erased shortly after their formation, typically within 10-30 Myr, or may not inflate at all following an energy injection. While this specific scenario is not visible in Figure~\ref{fig:subboxe_2063}, we do witness it occurring at other times within the cluster. 
    \item \textbf{Sequential X-ray cavity Inflation in the Similar Direction}: SMBH kinetic injection generally generates visible ripple structures in the cluster gas. Sequential energy injections can lead to interactions between these waves, further affecting the visibility and longevity of X-ray cavities.
\end{itemize}

These findings are consistent with observational studies, such as \cite{2014Bogdan_weak_shock,2021Ubertosi_slosh,2022Fabian_sloshing}, who suggest that interactions with cold fronts from sloshing motions or weak shocks can affect X-ray cavity expansion and lifetime. These processes introduce turbulence and varying pressure conditions in the ICM, potentially leading to the deformation or erasure of outer X-ray cavities before they can travel significant distances. Hydrodynamical simulations also suggest that turbulent ICM motions can disrupt X-ray cavities \citep[e.g.,][]{2012ApJ...750..166M,2019ApJ...871....6Y, 2020MNRAS.498.4983W}. However, it contrasts with the observational study of \cite{2023ApJ...954...56O} that found an absence of a correlation between X-ray cavity numbers and cluster dynamical state.

In TNG-Cluster, we can see X-ray cavities remaining visible for a few tens of Myr up to $\sim$200 Myr, with particularly stable ones identifiable at distances as far as 200 kpc from the SMBH. In contrast, most observational studies of X-ray cavities typically find them within 50 kpc of the SMBH, but with ages ranging from a few Myr to tens of Myr \citep[e.g.,][]{2004Birzan,2016Shin}. The greater distances of some TNG X-ray cavities suggest that they may move through the ICM at higher velocities, allowing them to cover larger distances over their lifetimes. However, there are notable observational exceptions that align more closely with these long-lived TNG X-ray cavities. For instance, \cite{2006Rafferty} report on the cluster MS 0735.6+7421, X-ray cavities with estimated ages of around 250 Myr and projected distances of approximately 170 kpc from the AGN core. 

\subsection{TNG X-ray cavities as a primary heating channel?}\label{subsec:coupling}

TNG-Cluster BCGs harboring X-ray cavities are associated with SMBHs in a low-accretion state (Figure~\ref{fig:fdbckmode}). In observational studies, X-ray cavities are commonly linked to AGNs slowly accreting below 10$^{-2}$ $\dot{M}_{\text{Edd}}$, where the energy brought by accretion is transformed into collimated winds and jets which are believed to be the main energy release channel \citep{2007McNamara}. In the TNG model, outbursts originating from the central SMBH emit energy at a rate ranging between \(10^{42}\) to \(10^{45}\) erg s$^{-1}$ (top panel of Figure~\ref{fig:subboxe_1gyr}). This energy release rate matches observations inferred from estimates of the amount of energy that AGN must release to explain the observed star formation rates in clusters. To effectively heat the ICM, this mechanical feedback must couple efficiently with the surrounding gas, transferring energy through X-ray cavities, shocks, and subsonic sound waves. The majority of observations show subsonic X-ray cavity inflation from AGN outbursts, with only rare cases of supersonic X-ray cavity expansion leading to shocks in the X-ray emitting gas \citep[e.g., NGC 4552,][]{2006ApJ...644..155M}. Although strong shocks may occur during the initial phase of X-ray cavity inflation, they likely account for a small fraction of the released energy into the ICM \citep{2022hxga.book....5H}. In TNG-Cluster, the majority of X-ray cavities either lack shocks or exhibit only weak shocks during their inflation stage, similar to observed X-ray cavities. Some TNG X-ray cavities feature bright rims with pressurized edges and weak shocks, reminiscent of cases observed in objects such as 4C+55.16 \citep{2022Timmerman} and MS0735.6+7321 \citep{2005McNamara}, where bright elliptical regions surrounding the X-ray cavities suggest the presence of an enveloping weak shock with a Mach number of $\sim1.4$. Observationally, it has been suggested that these weak shocks can deposit a significant amount of energy into the X-ray gas \citep[e.g.,][]{2015Randall,2023Ubertosi}. Additionally, the mixing of hot gas inside the X-ray cavity with the ICM may contribute to the heating budget. In TNG-Cluster simulations, X-ray cavities often exhibit a combination of weak shocks and inner gas that is ten times hotter. Evidence of mixing is seen in some of them, where a plume of hot gas trails behind the rising cavities  (third panel of Figure~\ref{fig:panel_cav_phys_pro}).

The role of sound waves generated by AGN feedback may also be significant. Prominent ripples are visible in the X-ray emission and UM images of TNG-Cluster X-ray cavities with a quasi-spherical morphology. A similar pattern is observed in the Perseus cluster \cite{2006MNRAS.366..417F}, where ripples, likely caused by sound waves, contribute to heating the cooling regions via dissipation and viscous damping. As can be seen in the outflow velocity measurements in Figure~\ref{fig:subboxe_panel} and \ref{fig:subboxe_1gyr}, as well as in the Mach number maps of the cluster core regions in Figure~\ref{fig:panel_cav_phys_pro}, there is an initial supersonic peak emanating from the SMBH at the time of energy injection (outflows $\sim$ 5-7,000 km/s, where the sound speed in the ICM is of the order of 3-4,000 km/s). The generated ripples are spherically propagating outward at greater distances and are characterized by weak (<2 Mach) shocks or no shocks. 

Understanding how this energy couples to the gas and the role of X-ray cavities in heating and stirring up the gas within the simulated TNG clusters remains a crucial question. Looking forward, exploring the role of X-ray cavities in these processes will provide insights into their broader impact on cluster evolution in the TNG cosmological framework.

\section{Summary and Outlook}\label{sec:conclusion}

In this work, we show that X-ray cavities can form naturally in simulated galaxy clusters as a result of SMBH-driven kinetic energy injections, i.e. an AGN mechanical feedback that is {\it not} implemented as relativistic, bipolar, collimated jets. For each of the 352 clusters in the TNG-Cluster cosmological simulation suite at $z=0$, a single mock Chandra observation was created from a random viewing angle. The mocks were systematically analyzed and visually inspected for X-ray cavities. We highlight the diverse morphologies and configurations of these X-ray cavities. The same implementation of SMBH feedback, combined with other components of the galaxy formation model and the hierarchical growth of structure within a cosmological context, yields a spectrum of diversity. This complexity is evident not only across different clusters, but also within individual clusters and within the same cluster over time. We summarize our major findings in the following points:

\begin{itemize}
    \item \textbf{Occurence}
    In the TNG-Cluster simulation, 39 per cent (136) of the 352 clusters exhibit X-ray cavities at $z=0$, visible in mock Chandra X-ray observations and enhanced by unsharp mask filtering.\\
    
    \item \textbf{Diversity and Morphologies}
    These structures display diverse morphologies and stages of evolution, from recently inflated X-ray cavities to detached and rising ones. We identify different configurations, including single, paired, and multiple X-ray cavities within the same cluster, and various shapes depending on the evolutionary stage (see the gallery in \S\ref{subsec:morpho} Figure~\ref{fig:panel_cav}). Single X-ray cavities are more common than pairs or multiples ( \S\ref{subsec:demogrpahics} upper panel Figure~\ref{fig:panel}). Unlike symmetric pairs often observed, TNG X-ray cavities show less symmetry, which may result from the model's kinetic feedback mechanism, where energy is injected unidirectionally into the gas around SMBHs, rather than the bipolar collimated outflows seen in some real galaxies.\\

    \item \textbf{Sizes} Manual measurements reveal X-ray cavity sizes ranging from a few to tens of kpc, with a mean radius of 9.2 kpc, and a weak positive correlation between X-ray cavity area and distance from the SMBH qualitatively consistent with observations (\S\ref{subsec:sizes}, Figures~\ref{fig:radius} and \ref{fig:area_distance}). \\
    
    \item \textbf{Thermodynamic Properties} Based on direct inspection of the cluster's core gas properties from the simulation—i.e. not inferred from the mocks— (see \S\ref{subsec:gas}, Figure~\ref{fig:panel_cav_phys_pro}),
    TNG-Cluster X-ray cavities are filled with gas of high temperature ($10^{7.8-8.1}$ K) and have lower densities compared to the surrounding ICM. 
    The gas inside is roughly in pressure equilibrium with the ICM. About 25 per cent of them exhibit X-ray bright and overpressurized dense edges, frequently coinciding with weak shock signatures. Conversely, most simulated X-ray cavities lack bright edges and typically do not show associated shocks.\\
    
    \item \textbf{Demographics of Hosting Halos}:
     Cool-core clusters in the TNG-Cluster simulation are more likely to harbor X-ray cavities, aligning with the scenario of a feedback loop driven by central cooling. Additionally, clusters with more X-ray cavities tend to be more dynamically relaxed, indicating that stable conditions favor the formation and/or longevity of these structures (see \S \ref{subsec:gas} the lower panels of Figure~\ref{fig:panel}). \\
     
    \item \textbf{Connection to AGN feedback}:
    TNG-Cluster X-ray cavities are a manifestation of episodic, kinetic, wind-like energy injections from SMBHs at the centers of clusters, which accrete at low Eddington ratios (see \S \ref{sec:origin} Figures~\ref{fig:subboxe_panel} and \ref{fig:subboxe_1gyr}).
    In the TNG model, which is used unchanged in TNG-Cluster, higher accretion rates -- indicating more luminous SMBHs -- lead to more powerful and frequent feedback, but this occurs within very low SMBH accretion regimes, typically with Eddington ratios $\sim10^{-3} - 10^{-5}$. X-ray cavities are inflated by bursts or momentum kicks given episodically in random directions by the central SMBH. This likely involves a push and weak shock mechanism driving the development of individual X-ray cavities.\\

    \item \textbf{Life cycle}: After an injection event, an X-ray cavity undergoes a rapid inflation stage, often accompanied by the formation of a weak shock front at the edges of the X-ray cavity. As the X-ray cavity detaches from the SMBH, it begins to rise within the ICM, expanding in size. Eventually, it dissipates and becomes indistinguishable in the ICM, or it may be disrupted by a sound wave resulting from a recent energy injection event (see one example in Figure~\ref{fig:subboxe_2063}). Their lifespans range from tens of Myr up to several hundred of Myr. Furthermore, clusters that contain multiple X-ray cavities are more likely to have formed due to successive feedback events, rather than from disruptions caused by turbulent motions in the ICM.
    
\end{itemize}

With this work, we demonstrate that AGN feedback in the IllustrisTNG comprehensive cosmological model can create a diverse X-ray cavity population at the current cosmic epoch, as revealed by the extensive TNG-Cluster simulation suite. In fact, the impact of AGN-driven X-ray cavities remains one of the major outstanding questions in understanding the physics of clusters. The IllustrisTNG model has demonstrated its success in effectively quenching star formation in massive galaxies via SMBH kinetic feedback at low accretion rates. X-ray cavities, as a direct outcome of mechanical feedback processes in these simulations, likely play a huge role in such a regulation. This paper sets up the basis for future investigations, establishing TNG-Cluster as a plausible framework for the formation of X-ray cavities within the context of a detailed galaxy formation model. This not only provides a new perspective for studying AGN feedback processes in clusters but also offers an additional valuable tool for evaluating the realism of cosmological simulations at galaxy cluster scales and in a spatially-resolved manner. 

On the one hand, TNG-Cluster opens up new avenues for studying the role of X-ray cavities in the overall feedback cycle of clusters, providing a deeper understanding of how these structures may influence and regulate the ICM and cluster evolution. Future analyses will be devoted to establishing the specific contribution of X-ray cavities to the overall heating of the ICM, to understanding the causal connection among X-ray cavities, regulation of star formation, and chemical enrichment of the ICM, to investigating the longevity of hereby simulated X-ray cavities and possible stabilization mechanisms, and the overall role of magnetic fields.

On the other hand, we can leverage X-ray cavities as an additional validation metric, to then further refine galaxy formation models. In fact, various AGN feedback prescriptions seem to be able to produce X-ray cavities, i.e. not only the IllustrisTNG model as demonstrated here but also the SIMBA implementation as recently shown by \cite{2024Jennings} at the galaxy group mass-scale in addition to the higher-resolution jet models of idealized cluster hydrodynamical simulations. Therefore, the mere existence of X-ray cavities in clusters is, yes, a strong and positive confirmation of the simulations' outcome but per se not necessarily a strong constraint on the model choices therein. A more quantitative and rigorous comparison between {\it populations} of observed and simulated X-ray cavities is required: this is the subject of a forthcoming companion paper (Prunier et al. in prep) where we conduct an apples-to-apples comparison between TNG-Cluster mocks and Chandra data.

\section*{Data Availability}
The IllustrisTNG simulations are publicly available and accessible at \url{www.tng-project.org/data}, as described in \cite{2019NelsonPublicReleaseTNG}, where the TNG-Cluster simulation data will also be made public in 2024-2025. All mock Chandra images generated for this paper will be made publicly available upon the release of TNG-Cluster. Other data directly related to this publication are available on request from the corresponding author.

\small{\section*{Acknowledgements}
MP thanks Urmila Chadayammuri, Florian Dedieu, Martin Fournier, Carter Rhea, Eric Rohr, and Tomáš Plšek for insightful discussions, as well as Nicolas Esser for comments on the manuscript. MP acknowledges funding from the Physics department of the University of Montreal (UdeM) and the Centre for Research in Astrophysics of Quebec (CRAQ). AP acknowledges funding from the European Union (ERC, COSMIC-KEY, 101087822, PI: Pillepich). JHL acknowledges funding from the Canada Research Chairs and from the Discovery grant program from the Natural Sciences and Engineering Research Council of Canada (NSERC). KL acknowledges funding from the Hector Fellow Academy through a Research Career Development Award. DN acknowledges funding from the Deutsche Forschungsgemeinschaft
(DFG) through an Emmy Noether Research Group (grant number NE 2441/1-1). 

The TNG-Cluster simulation suite has been executed on several machines: with compute time awarded under the TNG-Cluster project on the HoreKa supercomputer, funded by the Ministry of Science, Research and the Arts Baden-Württemberg and by the Federal Ministry of Education and Research; the bwForCluster Helix supercomputer, supported by the state of Baden-Württemberg through bwHPC and the German Research Foundation (DFG) through grant INST 35/1597-1 FUGG; the Vera cluster of the Max Planck Institute for Astronomy (MPIA), as well as the Cobra and Raven clusters, all three operated by the Max Planck Computational Data Facility (MPCDF); and the BinAC cluster, supported by the High Performance and Cloud Computing Group at the Zentrum für Datenverarbeitung of the University of Tübingen, the state of Baden-Württemberg through bwHPC and the German Research Foundation (DFG) through grant no INST 37/935-1 FUGG. 

All the analysis and computations associated to this paper have been realized on the Vera cluster of the MPCDF.}

\bibliographystyle{mnras}
\bibliography{TNG-Cav-Xray}

\appendix

\section{Gas cells sizes in the core regions of TNG-Cluster systems}

\begin{figure}
\centering
\includegraphics[width=0.4\textwidth]{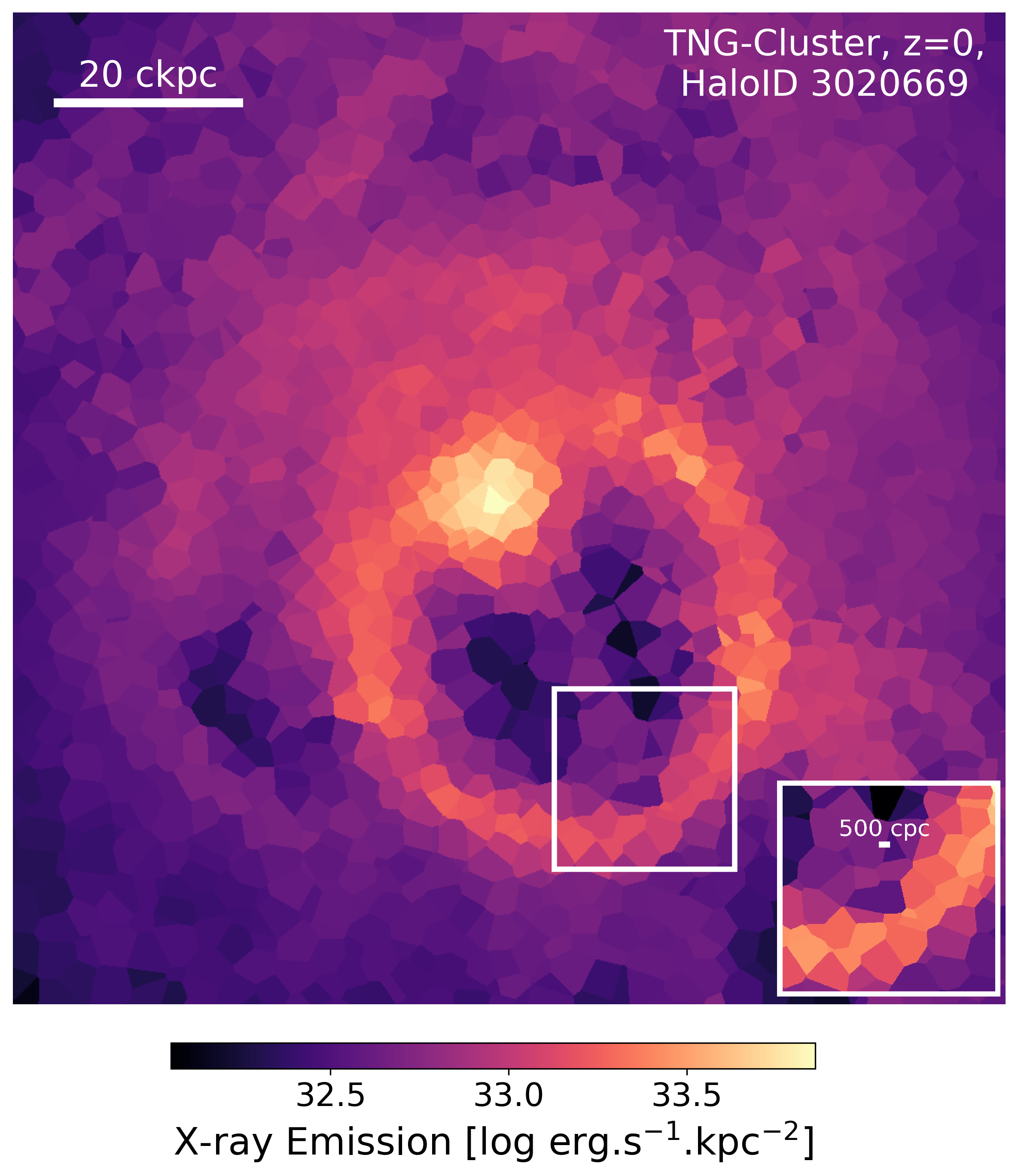}
  \caption{TNG-Cluster ID 3020669 at $z = 0$ across 120 ckpc a side, showing an inflating X-ray cavity, color-coded with X-ray emission. The maps depict the cross-section of the Voronoi tessellation of the gas, with a plane passing through the cluster center. Due to the quasi-Lagrangian nature of the AREPO code and the fixed target gas cell mass, smaller gas cells are resolved at higher densities.}
  \label{fig:annex_resolution_voronoi}
\end{figure}

\begin{figure}
\centering
\includegraphics[width=0.5\textwidth]{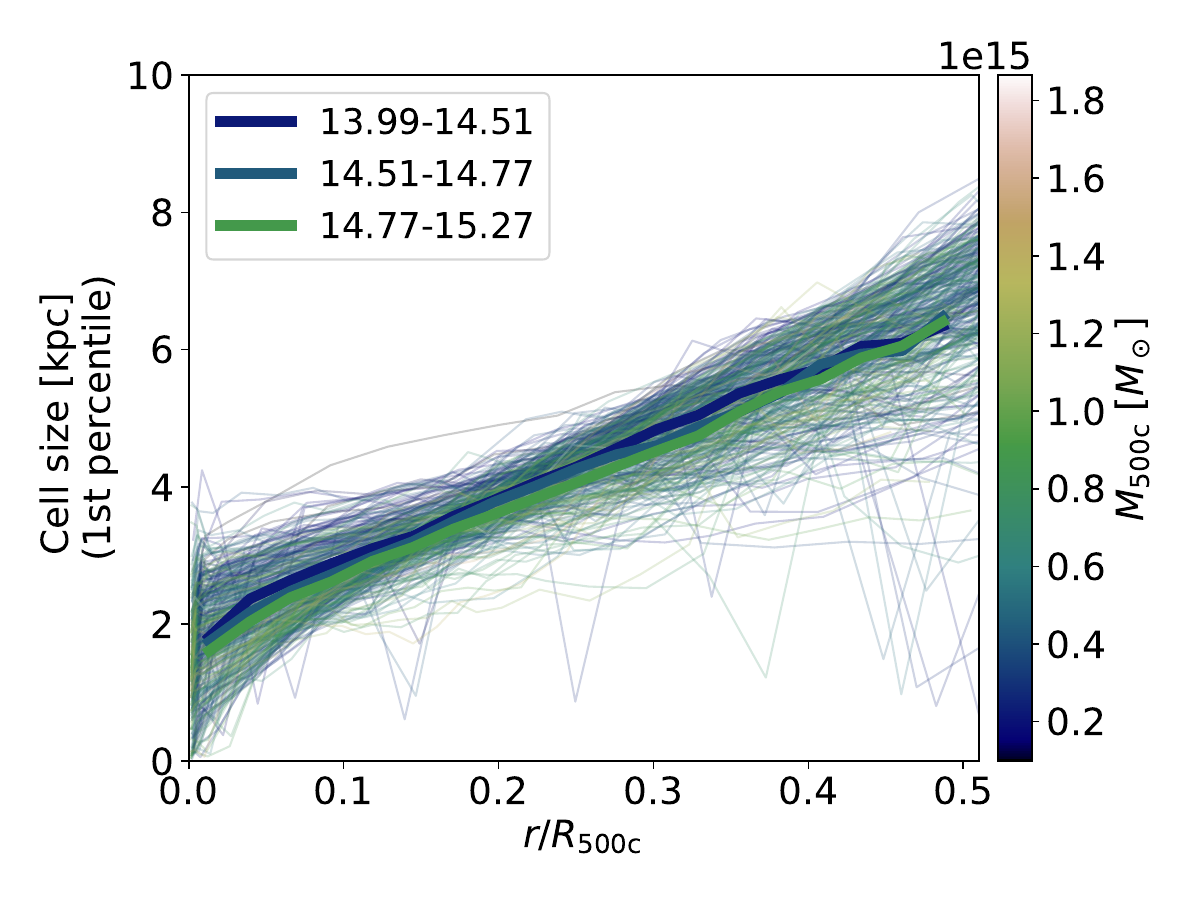}
  \caption{Radial profile of the gas cell sizes (1st percentiles of each cluster cell-size distribution) for 352 clusters in the simulation, color-coded by their mass M$_\text{500c}$. The three wider lines indicate the mean 1st percentile of gas cell size in clusters of a certain mass, as shown in the legend.}
  \label{fig:annex_resolution_radial}
\end{figure}

In the TNG-Cluster simulation, which shares the same resolution as TNG300 and employs the IllustrisTNG galaxy formation and evolution model, the equations governing gravity and magneto-hydrodynamics in an expanding Universe are numerically solved using the AREPO code \citep{2010Springel}. AREPO gas fluid dynamics are solved on an adaptive Voronoi-based moving mesh, that is locally density-dependent: smaller cells sample higher-density regions, while (de)refinement maintains a nearly fixed gas cell mass. A visualization of the underlying Voronoi tessellation in a TNG-Cluster galaxy is shown in Fig.~\ref{fig:annex_resolution_voronoi}, the cells can reach sizes below 1 kpc in high-density regions (e.g. the overdense X-ray cavity edges). In the cluster core region (<0.5R$_\text{500}$), the average gas spatial resolution (Voronoi-cell size) ranges from 1 to 8 kpc, and the 1st percentile of gas cell sizes is <6 kpc (Fig.~\ref{fig:annex_resolution_radial}).

\vspace{1cm}

\bsp	
\label{lastpage}
\end{document}